%% file: main.tex
\documentclass[11pt]{article}
\input{setup}

\title{Network Threshold Games\footnote{Langtry: University of Bristol (email: \texttt{alastair.lagtry@birstol.ac.uk}). Taylor: Jesus College, University of Cambridge (email: \texttt{srt44@cam.ac.uk}). Zhang: Christ's College, University of Cambridge (email: \texttt{yz558@cam.ac.uk}). We are grateful to George Charlson, Matt Elliott and Sanjeev Goyal for helpful comments. This work was supported by the Economic and Social Research Council [award reference ES/J500033/1]. Any remaining errors are the sole responsibility of the authors. }}

\author{Alastair Langtry \and Sarah Taylor \and Yifan Zhang}

\date{\today} 
\begin{document}
\maketitle
\begin{abstract}
\input{0.abstract}  
\emph{JEL Codes:} D85, O33
\emph{Keywords:} networks, threshold games, strategic complements, contagion, diffusion, coordination \\ 
\end{abstract}

\maketitle
\newpage
\input{1.text}
\newpage
\singlespacing
\bibliographystyle{abbrvnat}
\addcontentsline{toc}{section}{References}
\bibliography{bib}
\cleardoublepage
\onehalfspacing
\appendix
\numberwithin{equation}{section}
\numberwithin{thm}{section}
\numberwithin{prop}{section}
\numberwithin{defn}{section}
\numberwithin{rem}{section}
\numberwithin{lem}{section}
\numberwithin{cor}{section}

\newpage
\section{Proofs}\label{sec:proofs}
\input{A.proofs}

\newpage
\begin{center}
    \section*{Online Appendix}
\end{center}
\input{B.Online_Appendix}
\end{document}

%% file: setup.tex
\usepackage{natbib}
\usepackage{amsmath}
\usepackage{amsthm}
\usepackage{amssymb}
\usepackage{algorithm}
\usepackage{algpseudocode}
\usepackage{graphicx}
\usepackage{sidecap}
\usepackage{subcaption}
\usepackage{caption}
\usepackage{comment}
\usepackage{float}
\usepackage{lipsum}
\usepackage{changepage}
\usepackage{tikz}
\usepackage{pgf, pgfplots}
\usepackage{tikz-network}
\pgfplotsset{compat=1.17} 
\usepackage{xcolor}
\usetikzlibrary{calc}
\usetikzlibrary{automata}
\usetikzlibrary{shapes}
\usetikzlibrary{shapes.geometric}
\usetikzlibrary{decorations.markings}
\usetikzlibrary{arrows.meta, bending, positioning}
\usepackage{setspace}
\usepackage{fullpage}
\usepackage{pdfpages}
\usepackage{import}
\usepackage{rotating}
\usepackage{hyperref}
\usepackage{cleveref} 
\usepackage{minitoc}
\usepackage{algpseudocode} 
\usepackage{todonotes}
\usepackage{xfrac}
\renewcommand\fbox{\fcolorbox{white}{white}}
       
\newtheorem*{thm*}{Theorem}
\newtheorem{prop}{Proposition}
\newtheorem*{prop*}{Proposition} 
\newtheorem{cor}{Corollary}     
\newtheorem{lem}{Lemma}
\newtheorem*{lem*}{Lemma}
\newtheorem{defn}{Definition}
\newtheorem*{defn*}{Definition}
\newtheorem{rem}{Remark} 
\newtheorem*{rem*}{Remark} 
 
\newtheorem*{ex*}{Example} 

\newtheorem*{ass*}{Assumption}

\usepackage{comment}
\definecolor{green}{HTML}{51A351}
\definecolor{blue}{HTML}{2F96b4}
\definecolor{gray}{HTML}{BCBCBC}
\definecolor{gray1}{rgb}{0.62,0.65,0.65} 
\definecolor{gray2}{rgb}{0.29,0.31,0.31} 
\definecolor{gray3}{rgb}{0.06,0.07,0.08} 
\definecolor{teal1}{rgb}{0.13,0.54,0.52} 
\definecolor{teal2}{rgb}{0.18,0.34,0.32} 
\definecolor{brown1}{rgb}{0.59,0.53,0.49} 
\definecolor{brown2}{rgb}{0.81,0.73,0.64} 
\definecolor{yellow1}{RGB}{228,182,19} 
\definecolor{blue_low}{RGB}{137,187,255} 
\definecolor{blue_high}{RGB}{74,141,255} 
\definecolor{red_low}{RGB}{255,123,123} 
\definecolor{red_high}{RGB}{255,0,0} 

\newcommand{\monthyeardate}{\ifcase \month \or January\or February\or March\or April\or May \or June\or July\or August\or September\or October\or November\or December\fi, \number \year}

%% file: 0.abstract.tex
This paper proposes a new lens for studying threshold games played on networks when the thresholds are heterogeneous. These are games where agents have two possible actions, and prefer action $1$ if and only if enough of their neighbours choose action $1$. We propose a transformation of the network that `absorbs' the heterogeneity in thresholds into the network. This allows us to characterise equilibria in terms of the $k$-core -- a well-studied measure of network cohesiveness -- of the transformed network.
Our model is also the direct analogy to the workhorse model of \cite{ballester2006} when actions are $0$ or $1$. Further, our binary action version exhibits a remarkable stability property. When agents have linear-quadratic preferences, the $k$-core of the transformed network characterises the unique subgame perfect equilibrium of a sequential move version of the game -- no matter what order agents move in. 

%% file: 1.text.tex
At their core, threshold games on networks are simple: each agent has two possible actions (often called 0 and 1 for convenience), and prefers to take action 1 if and only if `enough' of her neighbours in the network do so too. The threshold quantifies `enough'. These games have been used to study a wide variety of different phenomena, including the adoption of technologies, uptake of government programs, participation in social behaviours and norms, and political protest.\footnote{See for example, \cite{morris2000contagion, reich2023diffusion} (technology), \cite{alexander2022algorithms} (program uptake), \cite{granovetter1978threshold, schelling2006micromotives, gagnon2017networks} (social behaviours and norms), \cite{chwe1999structure, chwe2000communication} (political protest).} 

In order to gain traction, existing work has typically required some simplifying assumption -- either on the thresholds or on the network. Simplifying the thresholds follows one of two routes. The first, following \cite{morris2000contagion}, assumes thresholds are a common fraction of the number of neighbours. The second, more popular in the computer science literature (e.g. \cite{seidman1983network, kong2019k, malliaros2020core}, has been to assume that all thresholds are a common number \citep{gagnon2017networks,harkins2013network}. These present different ways of abstracting away from the heterogeneity not coming from the network. 
Simplifying the network often involves assuming the agents are in some sense `homogeneous' in terms of network position \citep{jackson2006diffusion}, although earlier work often abstracted away from the network entirely and assumed that thresholds were in terms of behaviour of the whole population \citep{bass1969new, granovetter1978threshold, granovetter1986threshold}. Alternative approaches have varied the information structure, and found that incomplete information can allow for more tractable analysis \cite{galeotti2010network, leister2022social}.\footnote{\cite{galeotti2010network} consider a setting where agents only know their degree and the overall distribution of degree in the whole network. So agents can only condition their behaviour on their degree, which substantially simplifies analysis. \cite{leister2022social} assume that agents are uncertain about their thresholds.} 

This paper proposes an alternative approach. It assumes complete information, but avoids having to simplify either the threshold or the network. It does so by transforming the network in a way that `absorbs' the heterogeneity in the thresholds. Specifically, it gives all agents the same (absolute) threshold when we play the game on the transformed network. So by looking at the game through the lens of the transformed network, we gain traction without assuming away any of the heterogeneity.

To showcase the use of this new lens, we use it to provide some basic results about equilibrium behaviour in these games. First, we show that the maximal and minimal equilibria are characterised exactly by the $k$-core of the appropriately transformed network. A $k$-core is made up of the largest group of agents who are all tightly connected within that group, in the sense that they have at least $k$ neighbours in the group. It captures tightly `cohesive' parts of a network. Its link to the maximal equilibrium when agents all have the same threshold was established in \cite{gagnon2017networks}. Second, we use the transformed network and the $k$-core to characterise the other equilibria. 

Next, we show that when agents have linear-quadratic preferences a la \cite{ballester2006}, equilibrium behaviour exhibits remarkable stability with respect to the timing of the game. When agents move sequentially, there is a unique subgame perfect Nash equilibrium; characterised exactly by the $k$-core of the transformed network. Importantly, this remains the same not matter what order agents move in. With linear-quadratic preferences, our model also provides the direct analogy to the workhorse \cite{ballester2006} model in the case where actions are indivisible. Heterogeneity in the canonical linear-quadratic preferences demands heterogeneous thresholds when actions are indivisible. So previous work has been unable to provide the direct analogy.

By allowing for heterogeneous thresholds, our lens brings together the two existing approaches that have simplified the thresholds -- the common fractional threshold, and the common absolute threshold. Each approach has highlighted a measure of network cohesiveness as important for understanding equilibrium behaviour. The first highlights the $q$-cohesive set (which is a group of agents where everyone in the group has at least a fraction $q$ of their neighbours in that group), while the second approach highlights the $k$-core.
In light of this, we prove a tight equivalence between the $k$-core and the largest $q$-cohesive set. Viewed through our new lens, they are the same thing. This observation is in itself a new result. But it is also an important sanity check for our model. It ought to be the case that our object of interest -- the $k$-core of the transformed network -- works in both special cases.

Overall, we provide new tool for studying threshold games on networks, show how it can help characterise behaviour, and show how it relates to existing models. It could of course be used to help generalise existing application -- such as technology adoption or the spread of social norms -- by allowing for heterogeneity in agents' preferences or endowments. But it will also open up new applications where heterogeneity in the network structure and in agents' preferences/endowments are both essential features.

As one example, \cite{allmis2025social} study the welfare consequences of social mobility, show how this depends critically on the network structure. Here, heterogeneity in both agents' network position and endowments (and hence thresholds) are essential. Shutting down heterogeneity in thresholds would assume away any effects of social mobility -- the key focus of the paper. 
Using our transformation is therefore important for unlocking the analysis. Existing methods, relying on simplifying one dimension of heterogeneity, would be inappropriate here.

As another, \cite{ghiglino2024endogenous} study a model of peer effects where agents first select into groups, and then choose an action level (on the real line). They face social pressures to keep up with the action level of their neighbours -- but only those who selected into the same group. When all agents have the same ability, the group selection part of the model becomes a heterogeneous threshold game. Thresholds are heterogeneous despite this because the thresholds depend on both ability and degree in a novel way. Therefore our transformation would allow for a characterisation of who selects into which group. However, the focus in \cite{ghiglino2024endogenous} is on conditions under which all agents choose the same group. This amounts to providing conditions under which all agents choose action $1$ in the maximal or in the minimal equilibrium. As such, they do not need to use our transformation for that.

\subsection*{Related Literature}
Our paper contributes to the literature on threshold games on networks. In particular, it studies a complete information setting with an arbitrary fixed network and heterogeneous thresholds. The novel addition is tractably allowing for heterogeneous thresholds -- which reflect heterogeneity in agents' preferences or endowments -- while retaining a rich network structure. In doing so, it brings together the two related approaches that have been studied separately. The first assumes that thresholds are a common fraction of how many neighbours an agent has \citep{morris2000contagion, reich2023diffusion, jackson2025behavioral}, and often focuses more on coordination and diffusion of new behaviours. See \cite[Ch12.3]{goyal2023networks} for a textbook treatment. The second assumes that thresholds are a common absolute number \citep{gagnon2017networks, harkins2013network}. See \cite[Ch4.3]{goyal2023networks} for a textbook treatment. Our paper nests both approaches as special cases.\footnote{Due to the strategic complementarities, our model -- along with much of this literature -- belongs to the class of supermodular games. \cite{milgrom1990rationalizability, milgrom1994monotone} provide some foundational results for this more general class of games. Indeed, comparative statics of our model follow directly from \cite{milgrom1994monotone} (and hence we leave them to the Online Appendix).}

Closely related, \cite{acemoglu2011diffusion} study diffusion when agents have heterogeneous thresholds. But the focus is materially different. \cite{acemoglu2011diffusion} consider diffusion of action $1$ via myopic best responses from an initial set of seeds (agents who are assumed to take action $1$ before the game starts). So they only characterise what amounts to the minimal equilibrium in our model. Further, this initial set of seeds is random, and a key focus of their paper is to provide bounds on the expected number of agents who will take action $1$ given a certain number of randomly placed seeds.\footnote{The question of how to optimally choose the set of seeds in order to maximise the number of agents taking action $1$ has created its own strand of literature. It is often called a `seeding' problem. Important contributions here include \cite{kempe2003maximizing, kempe2005influential, banerjee2013diffusion, aral2013engineering, akbarpour2020just, jackson2025behavioral}. Our model has no seeds, and the question of targeting is out of scope here.}
In contrast, our setup is entirely deterministic, and we look at all equilibria.

Our paper also adds to the broader literature on network games.\footnote{\cite{jackson2015games, bramoulle2016games} and \cite{goyal2023networks} provide broad-based reviews.} 
Within this, there is a large body of work, following \cite{ballester2006}, focused on games with complete information and strategic complements. This work assumes that actions are continuous -- implicitly focusing on an intensive margin. We provide the exact analogy with binary actions. Taking these models, and restricting actions to be binary would yield exactly our model.\footnote{This applies equally to the highly tractable class of games with linear best responses (that follow \cite{ballester2006}), and to the much less tractable class with non-linear best responses. This is because when actions are binary, it is not meaningful to think of best-response as linear or non-linear -- they just have a threshold. In continuous action games, however, non-linearities in best responses present significant challenges. \cite{zenou2022network} provide a framework for solving those games.} 
This can be viewed as focusing instead on the extensive margin, or as studying the impact of introducing an indivisibility into the actions.
Binary actions have been studied elsewhere; but not with complete information. So they do not provide as close a comparison to the classic continuous action games. Notably, \cite{leister2022social} consider a setting where agents have imperfect information regarding their thresholds.

\section{Model}\label{sec:model}
\paragraph{Agents and the network.} There are $n$ agents, with typical agent $i \in N$. Agents are embedded in a weighted and directed network, represented by an $n \times n$ non-negative matrix $G$. In an abuse of terminology, we will often refer to $G$ as `the network'. 
We assume that the network is \emph{strongly connected}.\footnote{This means there is a directed \emph{path} from $i$ to $j$, for all $i,j \in N$. A directed path is a sequence of agents such that $G_{i,i'} >0$ for all $i,i'$ such that $i$ immediately precedes $i'$ in the sequence.}
The entry $G_{ij} \geq 0$ denotes the strength of a link from $i$ to $j$. An agent $j$ is $i$’s \emph{neighbour} if and only if $G_{ij} > 0$ (note that this need not be symmetric), and cannot be her own neighbour (so $G_{ii} = 0$ for all $i$). An agent's \emph{degree}, $d_{i}$, is the sum of her outward links, i.e. $d_{i} = \sum_{j \in N}G_{ij}$.\footnote{Formally, $d_i$ is a \emph{weighted out-degree}, but we will just call it degree for convenience.} 

\paragraph{Actions, Timing, Preferences.}
Each agent simultaneously chooses whether or not to take a binary action, $x_i \in \{0,1\}$. We assume that actions are \emph{strategic complements} and that preferences can be represented by a utility function $u_i(x_i, \sum_{j \in N} G_{ij} x_j)$ such that\footnote{In addition to strategic complementarities, \Cref{eq:prefs} embeds a mild smoothness property on the utility functions to ensure that the $k_i$'s exist. This is merely for clarity of exposition. We discuss this in Online \Cref{OA:relaxing_utility}.}  
\begin{align}\label{eq:prefs}
u_i\left(1, \sum_{j \in N} G_{ij} x_j \right) \geq u_i \left(0, \sum_{j \in N} G_{ij} x_j \right) \iff \sum_{j \in N} G_{ij} x_j \geq k_i \text{ for some } k_i \in \mathbb{R}.
\end{align}

\paragraph{Information and Equilibrium.}
We assume that preferences and the network structure are common knowledge and study Nash Equilibrium. In an abuse of terminology, we refer to the pair $(G,\mathbf{k})$ as the \emph{game}. We say the game exhibits \emph{strictly positive spillovers} if $u_{i}(\cdot)$ is strictly increasing in its second argument for all $i$. 

\subsection{Discussion: some terminology \& a known result}\label{sec:model_discussion}
The model we have set out is a \emph{threshold game} on a network. Its defining feature is that agents prefer action $1$ if the number of their neighbours taking action $1$ is above some threshold, and they prefer action $0$ otherwise. A well-known property of these games is that (1) there always exists a minimal and a maximal Nash equilibrium (in the set inclusion sense), and (2) the set of Nash Equilibria of the game forms a complete lattice under set inclusion \citep{milgrom1990rationalizability, vives1990nash}. 

As we noted in the introduction, a common way to gain traction in existing work has been to impose additional structure on the thresholds. This is either by specialising to an \emph{absolute threshold} setting ($k_i = k$ for all $i$, as in \cite{gagnon2017networks}), or to a \emph{fractional threshold} setting ($k_i = q d_i$ for all $i$, as in \cite{morris2000contagion}). Throughout the rest of the paper we will use these terms, \emph{absolute} and \emph{fractional} thresholds, to refer to these cases where all agents have the same threshold (either in absolute or fractional terms, respectively). And we will refer to our more general setting as the \emph{heterogeneous thresholds}.
%

Beyond allowing for heterogeneity \emph{per se}, one feature of a heterogeneous threshold model is that it allows for both `never takers' and `always takers' (i.e. those who will never/always take action $1$, regardless of what their neighbours do). Unaugmented, fractional threshold models allow for neither. To get round this, those models typically have an exogenous set of `seeds' -- agents who are exogenously fixed as taking action $1$ (perhaps despite their preferences) -- creating some `always takers'. In contrast, absolute threshold models allow for `never takers', but not `always takers'.

\section{The Transformation}
To gain traction in our more general case, we transform the network in a way that `absorbs' the heterogeneity in thresholds into the network, and gives all agents an absolute threshold.\footnote{It turns out that it is not possible to create an analogous transformation to give all agents a common fractional threshold. We discuss this in Online \Cref{OA:transform_fractional}.} 
This will help us study behaviour more clearly, and also bridge the gap between absolute threshold and fractional threshold games. Starting with any network, $G$, and set of thresholds, $\mathbf{k}$, we define a transformed network, $H(G,\mathbf{k})$, as follows:

\begin{defn}[Transformed network, $H$]\label{defn:adjustment}
Consider a set of agents $N$ embedded on a network $G$.

\begin{enumerate}
    \item Create a non-strategic ``shadow'' agent, denoted $s$, who plays $x_s = 1$, and let $H_{si} = \eta$ for all $i$, some $\eta \geq 1$,

    \item for $i$ with $k_i > 0$: let $H_{ij} = \frac{G_{ij}}{|k_i|}$ for all $j \in N$, and $H_{is} = 0$,

    \item for $i$ with $k_i < 0$: let $H_{ij} = \frac{G_{ij}}{|k_i|}$ for all $j \in N$, and $H_{is} = 2$,
    
    \item  for $i$ with $k_i = 0$: let $H_{ij} = G_{ij}$ for all $j \in N$, and $H_{is} = 1$,
\end{enumerate}
\end{defn}
For clarity, we omit the arguments of $H(\cdot)$ in the definition above, and will do so in the rest of the paper except where doing so would create confusion. Note that $H$ is an $(n+1) \times (n+1)$ matrix due to addition of the shadow agent. 

Using this transformed network gives us a game where all agents have a threshold of one. That is, each agent weakly prefers action $1$ (to action $0$) if and only if at least \emph{one} neighbour (in the weighted sense) in the transformed network, $H$, takes action $1$. Importantly, this transformation of the network is simply a new way of looking at the same game. It does not change the strategic incentives.

\begin{lem}[Best Responses unchanged]\label{lem:adjusted_network}
The games $(H,\mathbf{1})$ and $(G,\mathbf{k})$ are best response equivalent.\footnote{Additionally, as agents only have two possible actions, best response equivalence immediately yields better response equivalence (see \cite{morris2004best}).} 
And best responses are given by:
\begin{align}\label{eq:BR_H}
BR_i(\mathbf{x}_{-i}) =
\begin{cases}
    1 &\text{ if } \quad \sum_{j \in N \cup \{s\}} H_{ij} x_j > 1  \\
    \{0,1\} &\text{ if } \quad \sum_{j  \in N \cup \{s\}} H_{ij} x_j = 1  \\
    0 &\text{ if } \quad \sum_{j  \in N \cup \{s\}} H_{ij} x_j < 1
\end{cases}
\end{align}
\end{lem}

Because the games $(H,\mathbf{1})$ and $(G,\mathbf{k})$ have the same best responses, they must also have the same (set of) equilibria. While the main aim of \Cref{lem:adjusted_network} is a sanity check to show that our transformation is not changing the game in some important way, it has another use. It allows us to divide up the space of threshold games into equivalence classes. This is because many different threshold games $(G, \mathbf{k})$ and $(G', \mathbf{k'})$ can yield the same transformed network, $H$. And then, by \Cref{lem:adjusted_network}, they must be best response equivalent. So all games $(G, \mathbf{k})$ that induce the same adjusted network $H$ form an equivalence class.

\begin{cor}[Equivalence Classes]\label{cor:equivalence_classes}
The games $(G, \mathbf{k})$ and $(G', \mathbf{k'})$ are best response equivalent if and only if $H(G, \mathbf{k}) = H(G', \mathbf{k'})$.
\end{cor}

\paragraph{Network Cohesiveness.}
After our transformation, all agents have a threshold of one (in the transformed network). So we have converted the game into one with an absolute threshold. And existing work shows that a measure of network cohesiveness called the \emph{$k$-core} is important in these settings \citep{gagnon2017networks}. Intuitively, the $k$-core is the largest group of agents that all have at least $k$ links (in the weighted sense) to others inside the group. Formally,

\begin{defn}[$k$-core \cite{bollobas1984evolution}]\label{defn:k-core}
Given some $k \in \mathbb{R}^+$, the $k$-core of a network $G$, denoted $C(G,k)$, is the set of nodes forming the largest subgraph of $G$ such that all agents in $C(G,k)$ have at least $k$ (weighted) links to other agents in $C(G,k)$. 
\end{defn}

It can also be calculated using a simple algorithm that successively removes agents who do not satisfy a degree threshold. Simply (1) identify agents who have degree less than $k$, and (2) delete them along with their inward and outward links. Repeat this on the remaining sub-network until step (1) identifies no agents. We provide a more formal exposition of this algorithm in \Cref{sec:proofs}.

In contrast, a measure called a $q$-cohesive set has been shown to be the relevant measure in games with fractional thresholds \citep{morris2000contagion, reich2023diffusion}. Intuitively, a $q$-cohesive set is a set of agents such that all agents have a \emph{fraction} at least $q$ of their links to others inside the set. Naturally, this notion generalises to allow for heterogeneous values of $q_i$. In order to state this precisely, we let $\mathbf{k}(G,\mathbf{q})$ denote the absolute thresholds corresponding to fractional thresholds $\mathbf{q}$ and network $G$, i.e. $k_{i}(G,q) = q_{i}\sum_{j \ne i}G_{ij}$. Then formally,

\begin{defn} \label{def:q_cohesive_hetero}
    A set of agents $M \subseteq N$ in network $G$ is $\mathbf{q}$-cohesive if for each node in the set, the proportion of its neighbours also in $M$ is at least $q_{i}$. That is, if for all $i \in M$, $\sum_{j \in M} G_{ij} \geq q_i d_i$. 
\end{defn}

Notice that the $k$-core is the \emph{largest} set of agents that meets its cohesiveness criterion -- and hence it is unique. While in contrast, a $q$-cohesive set is \emph{any} set of agents that meets its cohesiveness criterion -- so there can be many such sets. So any relationship could only be between the $k$-core and the \emph{largest} $q$-cohesive set.

\paragraph{An equivalence.} Given that our model nests both the absolute and fractional threshold settings as special cases, there ought to be some relationship between the two definitions of network cohesiveness. It turns out that there is a relationship. And it is tight. For any fractional thresholds, $\mathbf{q}$, the largest $\mathbf{q}$-cohesive set of the network $G$ \emph{is} the $1$-core of the transformed network $H$. Formally,

\begin{prop} \label{prop:cohesive_core_equiv_hetero}
The maximal $\mathbf{q}$-cohesive set of the network $G$ is the $1$-core of $H(G,\mathbf{k}(G,\mathbf{q}))$.
\end{prop}

Despite their similar definitions, we are not aware of any formal link to date between $k$-cores and $q$-cohesive sets. \Cref{prop:cohesive_core_equiv_hetero} provides a very tight relationship. Additionally, this result shows the usefulness of allowing for heterogeneous thresholds. Even when $q_i = q$ for all $i$, the corresponding thresholds in absolute terms will in general be heterogeneous. So we must accommodate heterogeneous thresholds to show the equivalence.

Now that we have set out our transformation of the game and our key measure of network cohesiveness, we turn to characterising equilibrium play.

\section{Characterising Equilibria}\label{sec:eqa}
We already know that there is a complete lattice of Nash Equilibria (see \Cref{sec:model_discussion}), complete with a minimal and maximal. In this section we will first look at these two extreme equilibria, and then turn to the non-extreme ones.

\subsection{Extreme Equilibria}
Having applied our transformation, the game has become one with a common absolute threshold of $1$. So we know that the $1$-core of $H$ identifies the maximal equilibrium \citep{gagnon2017networks}. We can also see it from the mechanics of our algorithm for calculating the $k$-core. The algorithm first knocks out those who would not take action $1$, even if all of their neighbours did. It then knocks out those who would not take action 1, even if all of their remaining neighbours did. And it repeats this process until no more agents get knocked out. This is identical to a myopic iteration of best responses, starting from the point where everyone takes action $1$ initially. It is clear that this is an equilibrium. And no agent knocked out in the process could choose action $1$ in any equilibrium -- so it is maximal.

It also turns out that the $k$-core also characterises the minimal equilibrium. This stems from the fact that actions are binary: so the minimal equilibrium in terms of the set of agents taking action $1$ must also be the maximal equilibrium in terms of agents taking action $0$ (i.e. not taking action $1$). So we can ‘flip’ the game, and use our toolkit to characterise the maximal equilibrium in terms of action $0$. 
Before we transformed the network, we had that agent $i$ prefers action $1$ whenever at least $k_i$ of her neighbours are taking action $1$. But this is exactly the same as saying that she prefers action $0$ when at least $\tilde{k}_i = d_i - k_i$ of her neighbours are taking action $0$. So we rewrite the utility function:
\begin{align}\label{eq:complementary_prefs}
u_i\left(0, \sum_{j \in N} G_{ij} (1 - x_j) \right) \geq u_i \left(1, \sum_{j \in N} G_{ij} (1 - x_j) \right) \iff \sum_{j \in N} G_{ij} (1 -x_j) \geq \tilde{k}_i.
\end{align}

This yields a \emph{complementary game} $(G, \tilde{\mathbf{k}})$, with thresholds $\tilde{k}_i = d_i - k_i$. We can find the maximal equilibrium of this complementary game in exactly the same way we did for the original game: apply the transformation set out in \Cref{defn:adjustment}, and then calculate the 1-core of the resulting network. This identifies the maximal set of agents who take action $0$. The complement of this set is, by definition, the minimal set of agents who take action $1$. So it solves for the minimal equilibrium of the original game.

Taken together, we can characterise the maximal and minimal equilibria in terms of the $1$-core of a transformed network.

\begin{prop}\label{prop:extreme_eqa}
(i) In the maximal Nash Equilibrium, $\mathbf{x^{MAX}}$, agent $i$ takes action 1 if and only if she is in the $1$-core of the network $H(G, \mathbf{k})$. Formally, for all $i \in N$: $x_i^{MAX} = 1 \iff i \in C(H,1)$. \\

\noindent (ii) In the minimal Nash Equilibrium, $\mathbf{x^{MIN}}$, agent $i$ takes action 1 if and only if she is \emph{NOT} in the $1$-core of the complementary network $\tilde{H} = H(G, \mathbf{\tilde{k}})$, where $\mathbf{\tilde{k}} = \mathbf{d} - \mathbf{k}$. Formally, for all $i \in N$: $x_i^{MIN} = 1 \iff i \notin C(\tilde{H},1)$.
\end{prop}

There are a number of reasons why the extremal equilibria are particularly appealing to focus on. Most obviously, they provide upper and lower bounds on actions. 

Second, they are the outcomes of myopic best response dynamics when all agents start off with the same action (we arrive at the maximal equilibrium when all agents start from action $1$, and arrive at the minimal equilibrium when all agents start from action $0$). Myopic best response dynamics are one natural benchmark for thinking about equilibrium selection. They require little information (agents only need to know their own preferences and their neighbours’ actions) and are cognitively simple. And having all agents start from the same action is an appropriate starting point in many settings -- especially when thinking about the uptake of newly available actions, or the abandonment of ubiquitous ones. 
Myopic best response dynamics are also the standard benchmark in models of complex contagion.\footnote{In a complex contagion process, an agent adopts an action when at least a certain number ($\ge 2$) of her neighbours do so. In contrast, in a simple contagion process an agent adopts an action when at least one neighbour does so. The behaviour of these processes is markedly different. There is a large literature on each: see \cite{guilbeault2018complex, pastor2015epidemic}, and \citet[esp. Ch.19 \& 21]{easley2010networks} for reviews. \cite{centola2007complex} argue that complex contagion models tend to be more appropriate for social behaviours.} 
So the minimal equilibrium of our game the direct comparison to these models.

Third, when the game exhibits strictly positive spillovers, we can also view the maximal and minimal equilibria as extremes of agents' ability to coordinate. When the game exhibits strictly positive spillovers, the maximal equilibrium both Pareto dominates all other equilibria and is unique Coalition-Proof Nash Equilibrium (a la \cite{bernheim1987coalition}). These properties can be viewed as ways of trying to avoid coordination failures, and select the `best' equilibrium. Pareto dominance yields the outcome a utilitarian social planner would want (if she is constrained to choosing among equilibria). Coalition-proofness yields the outcome when any group of agents can explicitly coordinate their actions. 
Conversely, the minimal equilibrium is the outcome when agents start at everyone taking action $0$, and then play myopic best responses -- leaving them with absolutely no ability to coordinate; either explicitly through some communication, or implicitly through some forward-looking beliefs.

\subsection{Non-extreme Equilibria}\label{sec:non-extreme eqa}
We can also use the $k$-core of the transformed network to characterise the non-extreme equilibria. In any equilibrium, all `active' agents (those taking action $1$) must have at least $1$ link in $H$, in the weighted sense, to other active agents. This follows directly from the best response function (\Cref{eq:BR_H}). But then this means the sub-network of $H$ containing only the active agents and links between them must form a $1$-core. And all `inactive' agents must have less than $1$ link in $H$, in the weighted sense, to active agents. Otherwise they would prefer to be active. This means that adding any one inactive agent to the sub-network of active agents must \emph{prevent} the resulting network from being a $1$-core.

To formalise this discussion, we first need a formal definition of an induced sub-network. Then we can characterise all equilibria of the game.

\begin{defn}\label{def:induced_subgraph}
    For any set of agents $M \subseteq N$, the sub-network induced by $M$ in the network $H$, $H[M]$, is the network with the set of nodes equal to $M$ and $H_{ij}[M]=H_{ij}$ for any $i,j \in M$.
\end{defn}

In this result, we abstract away from exact indifferences. This is only to keep the result clean. It plays no other role. We dispense with this feature in Online \Cref{OA:non_extreme_indiff}. Doing so requires heavier notation, but provides no additional insight. 

\begin{prop}\label{prop:other_eq}
    Suppose that no agent $i$ is exactly indifferent between actions $0$ and $1$ for all strategy profiles $\mathbf{x}_{-i}$. Then for any $M$, the profile $\{x_{i}=1 : i \in M\}$ is a Nash Equilibrium if and only if: 
    (i) all agents in $M$ are in the 1-core of $H[M]$, and 
    (ii) $i$ is not in the 1-core of $H[M \cup \{i\}]$ for any $i \in N \setminus M$.
\end{prop}

\Cref{prop:other_eq} helps us understand what an equilibrium `looks like' in terms of the structure of the network. It consists of a cohesive group of agents that is almost `cut off' from the rest. That is, all agents in that group must be sufficiently well connected to others in the group. And no agent outside of the group can be too strongly connected to the group -- the group must be separated by a `moat' that is not crossed by strong `bridges' from anyone outside of the group. 

While this provides some sense of what an equilibrium looks like in terms of the transformed network, $H$, it may still be difficult to find all equilibria. In general, threshold games can have many equilibria, and there could be many cohesive groups, with various combinations of them jointly forming the set of active agents in an equilibrium. Nevertheless, there are important special cases where we can find all equilibria and characterise them exactly.

\paragraph{Two special cases.} While we treat the network structure as exogenous throughout this paper, in reality it must have come from somewhere. One natural candidate is that it may have been formed endogenously as part of a process that we have abstracted away from. And \cite{sadler2021games} have shown that two classes of networks -- \emph{Nested Split Graphs} and \emph{Ordered Overlapping Cliques} Graphs -- emerge from a wide range of network formation games. So if the network \emph{is} arising from some formation game we have pushed into the background, then we might reasonably expect that it belongs to one of these two classes. 

Nested Split Graphs are strongly hierarchical networks where if an agent $i$ has more friends than $j$, then $i$'s friends form a superset of $j$'s. Ordered Overlapping Cliques Graphs consist of sets of cliques -- groups where everyone in a given group is neighbours with everyone else in that group -- arranged in an overlapping fashion. These overlaps mean that some agents belong to more than one clique. To clarify results, we will also focus our attention on \emph{chain-linked cliques} graphs -- a special case of Ordered Overlapping Cliques Graphs where no more than two cliques overlap and these overlaps are all of the same size. We also assume that each clique is sufficiently large, but this is merely to simplify notation (we discuss what happens if we relax this assumption in \Cref{sec:proofs}, immediately following the proof to the result).

Additionally, it will be useful here to abstract away from all heterogeneity except for network position. This is a common approach in this literature (e.g. \cite{morris2000contagion, reich2023diffusion, gagnon2017networks} and \cite{hiller2017peer}) and helps focus attention on the role of the network. Specifically, we will consider a setting where the network $G$ is unweighted (so links have weight zero or one) and undirected, and all agents have a common threshold $\alpha>1$. This means we have $H_{ij} = H_{ji} \in \{0,1/\alpha\}$ for all $i,j \in N$ and there are no links to the shadow agent. 
This also fits more closely with work on network formation, where the resulting network is typically unweighted \citep{sadler2021games,  hellmann2013existence, hellmann2021pairwise}. We also assume there are no indifferences -- this would be easy to relax, but would add some extra notation without adding insight.

We start with the Nested Split Graphs, and present the formal definition and result together.

\begin{defn}
A network $H$ is a Nested Split Graph if $d_j^H \geq d_i^H \implies H_{j\ell} \geq H_{i \ell}$ for all $\ell \neq i,j$ (where $d_i^H$ is $i$'s degree in the transformed network $H$).
\end{defn}

\begin{rem} \label{rem:NSG}
    Suppose $H$ is an unweighted Nested Split Graph, with no links to the shadow agent. WLOG, order agents in descending order of their degree (so $d_i^H \geq d_{i+1}^H$ for all $i$). 
    There exists at most one non-empty equilibrium. In the non-empty equilibrium, there exists an integer $t>1$, such that $x_i=1$ if $i<t$, and $x_i=0$ otherwise. \footnote{Additionally, there always exists an empty equilibrium: $x_i=0$ for all $i$.} 
\end{rem}

The key observation is that Nested Split Graphs never have `moats' between groups. The strong hierarchical nature of the network structure means that action $1$ will always spread down the hierarchy. This spread can only stop when it reaches agents who do not have enough connections to \emph{ever} take the action. So there is only one `non-empty' equilibrium. There is also an `empty' equilibrium, where nobody takes action $1$. But this is just an artefact of assuming that there are no links to the shadow agent (which, recalling \Cref{defn:adjustment}, is simply that all agents had a positive threshold before we transformed the game).

In contrast, a chain-linked cliques graph can have many `moats' between the different cliques. This means there will be far more equilibria, and their structure will depend critically on the size of the overlap between two cliques. The size of the overlap affects how strong the bridges are between different cliques. Again, we present the formal definition and result together.

\begin{defn}[Chain-linked Clique Graph]\label{def:chain_clique}
An unweighted network $H$ is a \emph{chain-linked clique graph} if there exist cliques $\Gamma_1,\; \Gamma_2,\;\dots,\; \Gamma_L$ satisfying:
(1) \(\bigcup_{\ell=1}^L \Gamma_\ell = N\), and each \(\Gamma_\ell \subset N\) is a clique, and (2) for all $\ell$, $|\Gamma_\ell \,\cap\, \Gamma_{\ell+j}|$ is: $ \geq \alpha$ if $j =0$, $=z$ if $j=1$, and $=0$ if $j\geq2$.
\end{defn}

It is also helpful to refer to the overlap between two adjacent cliques as an `overlap', which we denote by \(B_\ell := \Gamma_\ell \cap \Gamma_{\ell+1}\). Recall that $|B_\ell| = z$. It is also helpful to define some objects that will help with the equilibrium characterisation.

First, pick an index-set $J \;\subseteq\;\{1,2,\dots,L\}$  

\begin{itemize}
    \item \textbf{Cliques profile:} $x_i = 1 $ if $i \in \Gamma_\ell$ for some $\ell \in J$, and $x_i = 0$ otherwise.

    \item A \textbf{gap} in $J$ is a set of consecutive cliques $h+1,...,h+m$ with $m\geq2$ such that $h+1,...,h+m \notin J$ and $h, h+m+1 \in J$. A \textbf{bridge} in $J$ is the union of overlaps in a gap. Let $B(J)$ denote the set of bridges in $J$.

    \item For some $J$ such that $B(J)$ is non-empty, and some $\mathcal{B} \subseteq B(J) \setminus \emptyset$, a \textbf{bridged cliques profile} is: $x_i = 1$ if either $i \in \Gamma_\ell$ for some $\ell \in J$ or $i \in \mathcal{B}$, and $x_i = 0$ otherwise.

    \item A cliques profile is \textbf{well-spaced} if all gaps in $J$ contain at least 3 cliques. 
\end{itemize}

Now that we have defined these objects, we are ready to state the formal result. 

\begin{rem}\label{rem:OOC}
Suppose \(H\) is a chain-linked clique graph. 

\noindent\textbf{(1) Wide moats,} $z < \frac{1}{3} \left( \alpha + 1 \right)$. A profile $\mathbf{x}$ is an equilibrium if and only if it is a cliques profile.

\smallskip
\noindent\textbf{(2) Moderate moats,} $\frac{1}{3} \left(\alpha + 1 \right) \,< \, z < 0.5 \alpha$.
A profile $\mathbf{x}$ is an equilibrium if and only if it is \emph{either} a cliques profile \emph{or} a bridged cliques profile.

\smallskip
\noindent\textbf{(3) Narrow moats,} $0.5 \alpha < z < \alpha$. A profile $\mathbf{x}$ is an equilibrium if and only if it is a well-spaced cliques profile.

\smallskip
\noindent\textbf{(4) No moats,} $ z > \alpha$. There are two equilibria: $x_i = 1$ for all $i$, and $x_i = 0$ for all $i$.
\end{rem}

While this result is rather terse, the intuition is straightforward. When there are `wide moats' in the network (i.e. the overlap between adjacent cliques is small), then any collection of cliques can be active (i.e. take action $1$). This is because the moat is wide enough between cliques for them to act independently. When there are `moderate moats', this is still true, but we add additional equilibria. Whenever two active cliques are separated by at least two inactive cliques, then there is also an equilibrium where the `bridges' between the active cliques are also active.

In contrast, when there are `narrow moats', the set of equilibria shrinks. Each clique cannot act fully independently. Combinations of active cliques where two active are separated by either one or two inactive cliques are \emph{not} sustainable in equilibrium. This is because in these cases, there will be a group of agents who are connected to two different active `overlaps' (i.e. overlaps that belong to at least one active clique). The action must spread to these agents -- being connected to two active bridges is sufficient for an agent to want to take the action.

Finally, with `no moats' actions will always spread. If one clique is active, this is always spread to the adjacent clique(s). So there are only two equilibria: either everyone is active or nobody is.

These two examples have helped build intuition around what the non-extreme equilibria `look like'. With this in place, we now turn to the robustness of equilibrium behaviour with respect to the timing of the game. 

\section{Robustness with linear-quadratic preferences}\label{sec:robustness}
Our model makes the benchmark assumption that all agents move simultaneously. In this section we examine what happens when we vary the timing of the game. To gain traction here, we need some additional structure on preferences. We opt for the canonical linear-quadratic preferences a la \cite{ballester2006}. 
So now preferences are described by:
\begin{align}\label{eq:BCZ_prefs}
    u_i = a_i x_i - \frac{1}{2} c_i x_i^2 + \phi_i \sum_j G_{ij} x_i x_j, \text{  with  } x_i \in \{0,1\},
\end{align}
and $a_i \geq 0, c_i>0, \phi_i >0$ for all $i$. In our game, this gives thresholds:
\begin{align}\label{eq:bonacich_thresholds}
    k_i = \frac{1}{\phi_i} \left( \frac{c_i}{2} - a_i \right).
\end{align}
It is clear that with these preferences our model is then identical to the canonical linear-quadratic game of \cite{ballester2006} \emph{plus an indivisibility}. That is, plus a restriction that agents can only take binary actions: $x_i \in \{0,1\}$ (recall that agents choose actions continuously in the canonical model). So our model provides a direct analogy to the workhorse model of \cite{ballester2006} in the case with binary actions. We briefly discuss how this change modifies some of the classic intuitions at the end of this section.

With these preferences, we can now see what happens when we vary the order in which agents move. It turns out that when agents move sequentially -- \emph{in absolutely any order} -- there is a unique Subgame Perfect Nash Equilibrium, where agents take action $1$ if and only if they are part of the $1$-core of the transformed network $H$. 

\begin{prop} \label{prop:sequential_game_BCZ}
    Suppose preferences are described by \Cref{eq:BCZ_prefs} and that all agents break indifferences in favour of action $1$.

    If agents make decisions sequentially -- $i$ chooses $x_i \in \{0,1\}$ at period $\sigma(i)$, for any permutation $\sigma: N \to N$ -- there is a unique Subgame Perfect Nash Equilibrium where $x_i^* = 1 \iff i \in C(H,1)$.
\end{prop}

This result has two related insights. First, it shows a new dimension of the robustness of the equilibrium predictions in threshold games on networks. No matter how we shuffle around the order in which agents move, the (subgame perfect) equilibrium does not change. Moreover, the prediction in a sequential move setting is the same as one of the equilibria in the simultaneous move setting. This robustness is not present when actions are chosen continuously. There, agents who move earlier in the sequence raise their actions (above their Bonacich centrality -- which is the unique equilibrium when actions are continuous) because they account for the fact that their higher action in turn induces higher actions by those who move later in the sequence.\footnote{This is easy to see, even with only two agents. We show such an example in the Online \Cref{OA:examples}.}

Second, it provides a new rationale for focusing on the maximal equilibrium in our simultaneous move model: it is the unique outcome that survives as a (subgame perfect) equilibrium were we to shift to sequential moves. This suggests that the maximal equilibrium -- characterised by the $k$-core of the transformed network $H$ -- is more robust than the other equilibria.

It is also useful to notice that our assumption about breaking indifferences in a particular direction is not demanding. This is because generically there will be no indifferences. If any situation with indifferences, perturbing a parameter value by a draw from a continuous and atomless distribution will break the indifference with probability $1$.\footnote{This is because indifference is a knife-edge case that requires $a_i - \frac{1}{2} c_i + \phi_i \sum_{j} G_{ij} x_j = 0$.} 

But an important feature of this parametrisation is that it shuts down `pure utility spillovers'. That is, an agent $i$ does not benefit from anyone else's action unless she takes the action herself. Mechanically, this is embedded by having $x_i=0$ as one of the two available actions.\footnote{That the other action is $x_i=1$ is, in contrast, just a normalisation.} 
So when an agent $i$ does not take the action (i.e. chooses $x_i=0$), then her payoff is a constant -- independent of all other agents' choices. So this special case is less suitable for settings where the action has (local) public goods properties. Instead, it is a better fit for technology adoption choices (broadly construed), where agents need to use the technology themselves in order to benefit from others' use. 

\paragraph{The impact of introducing an indivisibility.} As we noted above, one view of our model is the workhorse model of \cite{ballester2006} plus an indivisibility. So here we briefly discuss how adding this indivisibility affects the classic intuitions from the work on linear-quadratic games with continuous actions. First, equilibrium play is still characterised by a centrality measure -- but by a different one. With continuous actions, equilibrium actions are directly proportional to \emph{Bonacich Centrality} \citep{bonacich1987power}. With the indivisibility, actions in the maximal equilibrium are characterised by a centrality measure called the \emph{peeling value} \citep{abello2013fixed}. The peeling value is simply the highest value of $k$ for which agent $i$ is in the $k$-core of the network $H$.\footnote{It is easy to show that $k$-cores nest for increasing values of $k$. That is, $i \in C(H,k) \implies i \in C(H,k')$ for all $k' < k$. Plus, for any finite network, there will always be a value of $k$ such that the $k$-core is empty. Together, these observations ensure that the peeling value is a well defined object. It follows immediately from \Cref{prop:extreme_eqa} that the peeling value behaviour in the maximal equilibrium. The peeling value is sometimes referred to as \emph{coreness} \citep{harkins2013network, rodrigues2018network, saxena2020centrality}.} 
So we can see the peeling value as a direct analogy to Bonacich centrality for the binary action case.

Second, it makes the propagation of shocks `lumpy' rather than smooth. This is because agents must change their action discretely or not at all. So if a shock hits agent $i$, it is possible that some agents close to $i$ in the network are unaffected, while others very far away can change their behaviour. And increasing the size of the shock could have a larger impact, or none at all. As a result of this `lumpy' propagation, comparative statics results are limited. Agents' actions and welfare are weakly increasing in the strength of all links, and weakly decreasing in all thresholds. These results cannot be strengthened. A given change in a link strength or a threshold could have no impact on equilibrium behaviour, or it can have an impact whose size is bounded only by the number of agents in the network.\footnote{We state these results formally, and provide an example of this very large impact in Online \Cref{OA:examples}. Because our game exhibits strategic complements, it is also possible to obtain these results directly from \citet[Theorem 13]{milgrom1994monotone}, rather than by thinking about the transformed network and its $1$-core.}

Third, the indivisibility can create big changes in who is the \emph{key player} -- the agent whose removal leads to the largest reduction in total equilibrium actions. Importantly, seemingly peripheral players can be the key player. This is because removing an agent on the `periphery' of the network can trigger a large cascade of others switching away from action $1$. So it is possible that the shift to binary actions changes the identity of the key player dramatically. We show an example of this in Online \Cref{OA:examples}.

\section{A Planner's Problem}
We have shown how our transformation of the network facilitates analysis of network threshold games. As a final exercise, we consider how a social planner would alter the network to achieve some desired behaviour. As a simple benchmark, we suppose that the planner's goal is to have all agents take the action. So here the action is socially desirable. This could be because it creates positive externalities, or for some reason outside of the model. The tool available to the planner is building links,\footnote{A priori, we could allow for the planner to delete links as well as build them. But if deleting links is costly, then she will never do so. This is because all agents' actions are weakly increasing in the strength of every link. See Online \Cref{OA:comp_statics} for the formal comparative static result for this.} 
and she wants to achieve her goal as cheaply as possible. We also assume that the cost of building links is heterogeneous -- some are cheaper than others -- and that there is a limit as to how strong any one link can be. 

Additionally, we assume that the planner is able to coordinate on her most preferred equilibrium. Given that she wants to have all agents take the action, it is clear that she strictly prefers the maximal equilibrium over all others. Formally, the planner's problem is:

\begin{defn}[Planner's Problem]
Let  $\overline{G}$ and $\kappa$ be real, positive $n \times n$ matrices. $\overline{G}_{ij}$ is the maximum amount by which a link in the initial network, $G_{ij}$, can be strengthened, and assume that $\sum_j \overline{G}_{ij}$ is large enough. $\kappa_{ij}$ is the cost of strengthening a link from $i$ to $j$, and assume that for every $i$, $\kappa_{ij}\neq \kappa_{ij'}$ for every $j$ and $j'$ $\in N$. \\

\noindent Given $G, \overline{G}, \kappa$, the planner's problem is: 
\begin{align}
\min_{\, G^{soc} \in \mathbb{R}_{\ge 0}^{\,n \times n}} 
    &\; \sum_{i=1}^{n} \sum_{j=1}^{n} \kappa_{ij}\,\left(G^{soc}_{ij} - G_{ij}\right) \\
\text{s.t. } \ 
    &x^{MAX}_{i} =1 \ \ \forall i, \qquad
    G_{ij} \leq G^{soc}_{ij}  \leq G_{ij} + \overline{G}_{ij} \ \ \forall (i,j) \in N \times N. \notag
\end{align}
\end{defn}

This planner's problem has a simple solution. The planner chooses an agent arbitrarily, and builds outward links up to the point where that agent's degree is equal to her initial threshold, $k_i$. She starts at the cheapest link, strengthening it until it hits its capacity constraint. Then she moves to the next cheapest link. This repeats until the agent's degree reaches her threshold. The planner does this for all agents (obviously if an agent starts with a degree above her threshold, then the planner does not build any links for that agent). This procedure will always produce a unique outcome -- and therefore a unique optimal network.

\begin{prop}\label{prop:planner_cost_min}
     The planner's problem has a unique solution. For each $i$ with $\sum_{j}G_{ij}\geq k_{i}$, the planner does nothing. For each $i$ with $\sum_{j}G_{ij}<k_{i}$, the planner allocates outward links from node $i$ to nodes $j$ in ascending order of cost $C_{ij}$, up to their capacity constraint, so that:    
     \begin{equation}
       G^{soc}_{ij} =   
           \begin{cases}
               \overline{G}_{ij} + G_{ij} & \text{ if } \kappa_{ij}<\tau_{i}, \\
               k_{i} - \sum_{l: \kappa_{il} < \tau_i}(\overline{G}_{il} + G_{il}) & \text{ if } \kappa_{ij}=\tau_{i}, \\
               G_{ij} &\text{ otherwise.}
           \end{cases}
         \end{equation} 

    where $$\tau_{i} = \inf \left\{ \tau \in \mathbb{R}_{\geq 0}: \sum_{j: \kappa_{ij} \leq \tau}\left(\overline{G}_{ij}+G_{ij}\right) \geq k_{i}\right\}.$$ 
         
    This solution is implementable in polynomial time. 
\end{prop}

The key observation here is that the planner needs the \emph{whole} network -- after we have applied our transformation -- to be the $1$-core. This amounts to building a network where the very first step of the algorithm that calculates the $1$-core (which we explained immediately following \Cref{defn:k-core}) identifies no agents -- so all agents have degree of at least $1$ in $H$. Which is just the same as ensuring that all agents with $k_i>0$ have degree at least $k_i$ in the network $G^{soc}$.\footnote{And so have degree of at least $1$ in the transformed network $H$. Recall that agents with $k_i\leq 0$ have a link to the shadow agent, and so must always have a degree of at least $1$ in the transformed network $H$.}
But the planner does not want to add any links beyond what is needed for that -- because extra links are costly. So she will never build links that take an agent's degree strictly above $k_i$.

\paragraph{Related Problems.} While our set-up leads to a simple -- and fast-to-compute -- solution, some natural variants on the problem become computationally difficult. Indeed, the problem above appears to be the only natural variants of a social planner's problem on a non-empty network whose computational complexity does not explode with the size of the network in question.

The first, and closest, variant is to consider a setting where the planner can only build undirected links and whose weight cannot be modified continuously. This corresponds to a setting where the planner has less fine-grained control over how to build the network, and can only create a relationship between two agents. While natural in some settings, this problem is NP-hard. This is because undirectedness means that each agent cannot be activated separately and also faces discreteness issues. It would therefore become an integer programming problem, which is known to be NP-hard \citep{karp2009reducibility}.

Another natural variant would be flip the objective and constraint, and suppose that the planner aims to maximise the number of active agents while facing some budget constraint. It turns out that this version of the problem is NP-hard. We set out this planner's problem formally, and prove the NP-hardness in Online \Cref{OA:planner_problem}. 

The headline logic behind that result is that maximising the number of agents who take the action involves leveraging chains of spillovers -- building a link to induce one agent to take the action can then induce a cascade of others doing so too. Each potential new link can create a different cascade. But importantly, each group of new links can create a different cascade -- and this need not be a straightforward function of the cascade from each link in the group. So the planner needs to evaluate the impact of each possible group of new links separately -- which quickly becomes a combinatorially difficult problem. 

This difficulty ceases to apply when (as in \Cref{prop:planner_cost_min}) the planner has enough resources to make all agents take the action because full activation is clearly optimal in this case. Hence, the planner does not need to make the computationally costly decision of which set of agents to activate. Picking the optimal (partial) set of agents to activate with a non-empty initial network 
requires the planner to calculate the cascade associated with activating every budget-feasible subset of initially inactive agents, and then compare the outcomes. The sensitivity of cascades to the initial network makes pruning candidates on some general criteria hard. Thus, the number of subsets to be checked grows exponentially in the number of initially inactive agents. 


\section{Conclusion}
In this paper, we provide a new way of looking at threshold games when the network is fixed and information is complete. We have shown how transforming the network in the right way `absorbs' the heterogeneity in thresholds, leaving a simpler game. We can then use the $k$-core -- a known measure of network centrality -- to characterise equilibria of the game. And when agents have linear-quadratic preferences, the $k$-core turns out to characterise the unique subgame perfect equilibrium of a sequential move version of the game -- no matter what order players move in. This shows that equilibrium behaviour is, in a sense, much more robust when actions are $0-1$ compared to when they are continuous. Recall that our threshold game is the direct analogy to the workhorse model of \cite{ballester2006} when we restrict actions to be binary. 

From here, there are two natural directions for future work. First, our approach will facilitate analysis in more applied settings where heterogeneity in both the network structure and in the thresholds are critical. Second, there is a lack of work on threshold games in a random network setting. With a fixed network, comparative statics are almost `anything goes' results. A random networks approach may permit much more powerful comparative statics results.

%% file: A.proofs.tex
\paragraph{Proof of \Cref{lem:adjusted_network}}
By assumption, $BR_i(\mathbf{x}_{-i}) = 1$ if and only if $\sum_j G_{ij} x_j \geq k_i$.

\noindent \textbf{Case A:} if $k_i > 0$, then $H_{is} = 0$ and $k_i = |k_i|$. So $\sum_j H_{ij} x_j = \sum_{j \neq s} \frac{G_{ij}}{k_i} x_j$. It is then immediate that $\sum_j H_{ij} x_j \geq 1$ is identical to $\sum_{j \neq s} G_{ij} x_j \geq k_i$.

\noindent \textbf{Case B:} if $k_i < 0$, then $H_{is} = 2$ and $k_i = -|k_i|$. 
So $\sum_j H_{ij} x_j = \sum_{j \neq s} \frac{G_{ij}}{|k_i|} x_j + 2 x_s \geq 1$. Hence $\sum_{j \neq s} G_{ij} x_j \geq (1 - 2 x_s) |k_i|$. Noting that $x_s = 1$, and hence $(1 - 2 x_s) |k_i| = k_i$, yields the result.

\noindent \textbf{Case C:} if $k_i = 0$, we have $H_{is} = 1$. So $\sum_j H_{ij} x_j = \sum_{j \neq s} G_{ij} x_j + x_s \geq 1$. This is identical to $\sum_{j \neq s} G_{ij} x_j \geq (1 - x_s)$. Noting that $x_s = 1$, and hence $(1 - x_s) = 0 = k_i$, yields the result. \hfill \qed

\paragraph{Proof of \Cref{cor:equivalence_classes}} follows directly from \Cref{lem:adjusted_network}. \hfill \qed

\paragraph{Proof of \Cref{prop:cohesive_core_equiv_hetero}}
Let $Q$ denote the maximal $\mathbf{q}$-cohesive set. We show that $Q$ coincides with the maximal equilibrium.\footnote{That a maximal equilibrium exists is a known result \citep{vives1990nash,jackson2008}. For completeness, we provide a simple proof in the Online Appendix.} 
By definition of $\mathbf{q}$-cohesive set, for all player $i \in Q$, if all other players in $Q$ choose action 1, $i$'s best response is action 1. Also, for all player $i \in N\setminus Q$, the proportion of its neighbours in $Q$ must be less than $q_{i}$, or otherwise $Q \cup i$ is also a $\mathbf{q}$-cohesive set. Therefore, if players in $Q$ choose action 1 and all other players choose action 0, then best response of players in $ N \setminus Q$ is action 0. Therefore, $Q$ is an NE. For any set $M \supset Q$, there exists some player $i \in M$ such that its fraction of neighbours in $M$ is less than $q_{i}$. Therefore, the strategy profile where players in $M$ choose action 1 and others choose action 0 can't be equilibrium. \hfill \qed \\ 


\noindent Before stating the proof to \Cref{prop:extreme_eqa}, we present an algorithm for calculating the $k$-core. This provides a formal statement of the procedure described following \Cref{defn:k-core}. 

\begin{algorithm}
\begin{algorithmic}
\State \textbf{Input}: Set of nodes in the network, $N$ and adjacency matrix for the network, $G$. 
\State \textbf{Input}: A threshold, $k$.
\State \textbf{Output}: Set of nodes in the $k$-core, $C^{k}$.
\State Set $C^{k} = N$

\Repeat:
\State $L = \{i: \sum_{j \in C^{k}}G_{ij} < k\}$
\State Set ${C^{k}}_{old} = C^{k}$
\State Set $C^{k} = C^{k} \setminus L$
\Until{${C^{k}}_{old} = C^{k}$}

\State \Return{$C^{k}$}
\end{algorithmic}
\caption{Calculating the $k$-core of a network $G$ \citep{gagnon2017networks}}\label{alg:k_core}
\end{algorithm}

\paragraph{Proof of \Cref{prop:extreme_eqa}}
\textbf{Part (i).}
First, it follows from \Cref{lem:adjusted_network} that the game $(H, \mathbf{1})$ is equivalent to the game $(G, \mathbf{k})$. Second, it is clear that, at each of its iterations, the algorithm in \Cref{defn:k-core} deletes the set of agents who do not have enough remaining neighbours to meet the threshold (of $1$) for taking action 1. 

Therefore when the algorithm terminates, the remaining agents are exactly those who have at least $1$ outward links (in the weighted sense) to others who have also not been deleted. This means they will take action 1 if all of their neighbours do so. Therefore the $1$-core is an equilibrium. Maximality follows from the fact that the algorithm only deletes agents who will \emph{never} take the action. Uniqueness follows from maximality. 

\textbf{Part (ii)} follows from part (i) coupled with the discussion of the complementary game that immediately precedes the statement of the Proposition (especially \Cref{eq:complementary_prefs}). \hfill \qed

\paragraph{Proof of \Cref{prop:other_eq}}
\textbf{Part 1: $H[A]$ is a $1$-core and $H[A \cup \{i\}]$ is \emph{not} a $1$-core $\implies$ $A$ is a Nash Equilibrium.} By the definition of the $1$-core, we have $\sum_{\ell \in A} H_{j \ell} \geq 1$ for all $j \in A$. Therefore all $j \in A$ prefer to choose $x_j = 1$ if all other $j \in A$ do so too (regardless of the choices of agents $i \notin A$). Also by the definition of the $1$-core, we have $\sum_{\ell \in A} H_{i \ell} < 1$ for all $i \notin A$.\footnote{We can see this by a simple contradiction argument. If it were not the case, then $A \cup \{i\}$ would constitute a $1$-core. Which would be a contradiction.} 
So all $i \notin A$ prefer to choose $x_i = 0$ if all other $i \notin A$ do so too (regardless of the choices of agents $j \in A$). Therefore $x_i = 1$ for all $i \in A$ and $x_i = 0$ for all $i \notin A$ is a Nash Equilibrium.

\textbf{Part 2: $A$ is a Nash Equilibrium $\implies$ $H[A]$ is a $1$-core and $H[A \cup \{i\}]$ is \emph{not} a $1$-core.} For $A$ to be a Nash Equilibrium, we must have: $\sum_{\ell} H_{j \ell} x_{\ell} \geq 1$ for all $j \in A$ and $\sum_{\ell} H_{i \ell} x_{\ell} \leq 1$ for all $i \notin A$. Additionally (and by definition), $x_j = 1$ for all $j \in A$ and $x_i = 0$ for all $i \notin A$. Substituting these into the inequalities yields: $\sum_{\ell \in A} H_{j \ell} \geq 1$ for all $j \in A$ and $\sum_{\ell \in A} H_{i \ell} \leq 1$ for all $i \notin A$. 

By assuming no indifferences, we have ruled out $\sum_{\ell \in A} H_{i \ell} = 1$ for any $i \notin A$. Finally, note that by definition having $\sum_{\ell \in A} H_{j \ell} \geq 1$ for all $j \in A$ means $H[A]$ is a $1$-core. And also by definition, having $\sum_{\ell \in A} H_{i \ell} < 1$ for all $i \notin A$ means that $H[A \cup \{i\}]$ is \emph{not} a $1$-core of $H$ for any $i \in N \setminus A$. \hfill\qed

\paragraph{Proof of \Cref{rem:NSG}}
We first show that if agent $i$ takes action 1, then all agents with indices smaller than $i$ must choose action 1. 
By construction, for any $j<i$, we have $d_{j} > d_{i}$. If action 1 is optimal for agent $i$, then at least 1 of $i$'s neighbours (in the weighted sense) chooses action 1. Since $G$ is a nested split graph, all neighbours of $i$ are also neighbours of $j$, and thus at least 1 of $j$'s neighbours choose action 1. This implies that agent $j$ must choose action 1. 
%
%
This shows that the equilibrium must take a 'cut-off' structure; i.e. there exists some $t$ such that agents with smaller indices choose 1 and those with weakly larger indices choose 0. 

We now show that there is at most one such 'cut-off' greater than 1. Suppose that there exist two such cut-offs. $t,t'>1$ with $t>t'$. Recall that each agent needs at least $\alpha$ neighbours to choose action 1 in order for them to choose action 1. So if \emph{any} agents are active, there must be at least $\lceil \alpha+1 \rceil$ active agents. Therefore we know that $t'>\alpha+1$. If all agents with indices weakly larger than $t'$ have fewer than $\alpha$ neighbours, they cannot choose action 1. This contradicts $t$ being a cut-off. If there exists some agent $i > t'$ with more than $\alpha$ neighbours, then by the definition of nested split graphs, $i$'s neighbours are the $d_{i}$ agents with highest degrees. In other words, $i$'s neighbours are agents $1,2,...d_{i}$. This means that $i$ must have at least $\alpha$ neighbours choosing $t$ in the candidate equilibrium associated with cut-off $t'$, and has incentive to deviate to action 1.     \hfill \qed

\paragraph{Proof of \Cref{rem:OOC}}
Equilibrium behaviour in chain-linked cliques satisfies the following property:
\begin{lem} \label{lem:OOC_actions}
In any equilibrium, the set of agents choosing action 1 is the union of a collection of cliques and overlaps.
\end{lem}
We prove this lemma in Online \Cref{OA:cliques_lemma}. We now return to the proof of \Cref{rem:OOC}. 

\paragraph{Case 1, Wide moats, $z < \frac{1}{3}(1+\alpha)$:} 
\textbf{Any cliques profile is an equilibrium:} In any such profile, if $x_{i}=1$, then all agents in the same clique as $i$ also choose 1. They all have at least $\alpha$ neighbours choosing action $1$, so they do not wish to deviate. If $x_{i}=0$, then $i$ must be in a gap. This implies that $i$'s neighbour must also be in the gap or an overlap at the ends of the gap. Therefore, $i$ can have at most $2z$ neighbours who are not in the gap and choose action 1. Since $2z<\alpha$, $i$ has no incentive to deviate.  

\textbf{No other profile can be an equilibrium:} By \Cref{lem:OOC_actions} point 2, any other profile must involve an overlap $B_{l}$ taking action 1 and $B_{l}$ is in a gap. By argument in the last paragraph, agents in $B_{l}$ have at most $2z$ neighbours who are not in $B_{l}$ and choose action 1. Therefore, agents in $B_{l}$ have at most $3z-1<\alpha$ neighbours choosing action 1, so they want to deviate to 0. 

\paragraph{Case 2, Moderate moats, $\frac{1}{3}(1+\alpha) < z < 0.5 \alpha $} \textbf{Any such profile is an equilibrium: } If $x_{i}=1$ and all agents in $i$'s clique choose 1, then $i$ has no incentive to deviate as in case 1. If $x_{i}=1$ and $i$ is in a bridge, then $i$ must be in an overlap and both of $i$'s adjacent overlaps must choose action 1. Therefore, $i$ has at least $3z-1>\alpha$ neighbours choosing action 1. 

If $x_{i}=0$ and $i$ is in a gap whose associated bridge does not choose action 1, then the same proof as in case 1 shows that $i$ does not want to deviate. Otherwise, $i$ must be in some gap and not in any overlap. Therefore, $i$ has $2z<\alpha$ neighbours choosing action 1, and does not want to deviate. 

\textbf{No other profile can be equilibrium:} By \Cref{lem:OOC_actions}, any equilibrium can be represented as a union of cliques and overlaps. We can find the gaps as defined in the main text for the set of cliques in any candidate equilibrium. If the candidate profile is not a cliques profile or bridged cliques profile, then there must exist some gap in which some overlap $B_{l}$ chooses action 1 and at least one of $B_{l-1}$ and $B_{l+1}$ chooses action 0. Therefore, agents in $B_{l}$ have at most $2z<\alpha$ neighbours choosing action 1, and they want to deviate. 

\paragraph{Case 3, Narrow moats, $0.5 \alpha <  z < \alpha$}: \textbf{Any such profile is an equilibrium:} Agents choosing action 1 do not want to deviate for the same reason as in case 1. In a well-spaced cliques profile, if an agent $i$ has neighbours in the overlap at one end of the gap, then $i$ cannot have neighbours in the overlap at the other end. Therefore, agents in any gap have at most $z<\alpha$ neighbours choosing action 1, and they have no incentive to deviate. 

\textbf{Any other profile cannot be equilibrium:} If agents in $B_{l}$ choose action 1, it cannot be the case that all others in $\Gamma_{l}$ and $\Gamma_{l+1}$ choose action 0, since $z<\alpha$. Therefore, by \Cref{lem:OOC_actions}, agents in either $B_{l+1}$ or $B_{l-1}$ must choose action 1. Since $2z>\alpha$, agents between $B_{l}$ and the adjacent overlap that chooses action 1 must also choose action 1. The argument shows that there can be no gap in which some overlaps choose action 1, and thus any equilibrium must be a cliques profile. If the cliques profile is not well spaced, then there exists a gap which contains only 1 overlap $B_{l}$. Agents in overlaps at both ends of this gap are neighbours of $B_{l}$, so agents in $B_{l}$ want to deviate to action 1. 

\paragraph{Case 4, No moats, $z > \alpha$:} It is trivial to verify that all agents choosing 0 and all agents choosing 1 are equilibria. If some agent $i$ chooses action 1, then by \Cref{lem:OOC_actions}, all agents in the clique $\Gamma_{l}$ that contains $i$ must choose action 1. Agents in adjacent cliques have $z>\alpha$ neighbours in $\Gamma_{l}$, so they must also choose action 1. Applying this iteratively shows that if $x_{i}=1$ for some $i$ in an equilibrium, then all agents must choose 1. \hfill \qed

\paragraph{Proof of \Cref{prop:sequential_game_BCZ}}
We prove this by induction on the number of agents. The claim holds trivially for all games with one agent. Suppose now that the maximal equilibrium of the simultaneous move game is the unique SPNE outcome for all sequential games with $n$ agents. In an abuse of terminology, we will use `maximal equilibrium' to mean `maximal equilibrium of the simultaneous move game'.

Suppose we have a sequential move game with $n+1$ agents, who are labelled according to their sequence of action. Fixing the first agent's action defines an $n$-agent continuation game. Thus there are two $n$-agent continuation games: one for when agent 1 is fixed to take action 1 and another when agent 1 is fixed to take action 0. 
By the induction hypothesis, each of these $n$-agent continuation games has a unique SPNE outcome, which is their respective maximal equilibria.\footnote{Notice that the $n$ agent continuation game is not the same as a game with only $n$-agents. In the continuation game where agent 1 is fixed to action 0, the continuation game is equivalent to removing agent 1 from the network. In the continuation game where agent 1 is fixed to play action 1, the continuation game is not equivalent to simply removing agent 1 from the network. Rather, it is equivalent to game on a network of $n+1$ agents where agent 1 plays 1 non-strategically and the remaining $n$ agents play strategically.}

Let $\mathbf{x^{c}}(x_{1})$ be the maximal equilibrium which is played in the $n$-agent continuation game if agent 1 chooses action $x_{1}$. Furthermore, let  $\mathbf{x}^{MAX}_{-1}$ be the profile of actions played by all agents excluding agent 1 in the maximal equilibrium of the $n+1$ player game.\footnote{In other words, this is the tuple which consists of the actions of players $\{2, \ldots, n+1\}$ in $\mathbf{x}^{MAX}$.} 

We first note that $\mathbf{x^{c}}(x_{1}^{MAX}) = \mathbf{x}^{MAX}_{-1}$. In $\mathbf{x}^{MAX}_{-1}$, agents must be playing best responses to each other and to agent 1 playing $x_{1}^{MAX}$. Therefore, $\mathbf{x}^{MAX}_{-1}$ must also be an equilibrium given agent 1 is playing $x_{1}^{MAX}$. By maximality of $\mathbf{x}^{MAX}$, $\mathbf{x}^{MAX}_{-1}$ must be the maximal equilibrium given that agent 1 plays $x_{1}^{MAX}$, which is $\mathbf{x^{c}}(x_{1}^{MAX})$ by definition. This implies that $\mathbf{x}^{MAX} = (x_{1}^{MAX},\mathbf{x^{c}}(x_{1}^{MAX}))$. Given this, it's sufficient to show that agent 1 chooses $x_{1}^{MAX}$. We consider two cases:

\emph{Case 1}: Suppose that $x^{MAX}_{1}=0$. If first mover prefers to choose action 1, agents choosing action 1 in the profile $(x_{1}=1, \mathbf{x^{c}}(1) )$ get positive payoff. Therefore, their best response to $(x_{1}=1, \mathbf{x^{c}}(1) )$ is action 1. By complementarity of actions, there must exist an equilibrium where agent 1 chooses action 1, contradicting maximality of $\mathbf{x}^{MAX}$.

\emph{Case 2}: Suppose that $x^{MAX}_{1}=1$. Agent 1 must get weakly positive payoff from choosing action 1, and thus a deviation to action 0 can't be profitable. \hfill \qed

\paragraph{Proof of \Cref{prop:planner_cost_min}}
Only agents $i$ with strictly positive thresholds are candidates for $G^{soc}_{ij} > G_{ij}$ for some $j$, as all other agents automatically have $x^{MAX}_{i}=1$. Moreover, all agents with $\sum_{j}G_{ij}/k_{i} \geq 1$ will receive no intervention. This is because when we calculate the $1$ core of $H$, we begin by deleting all nodes with out-degree $<1$. If the degree of these nodes is increased to above $1$, then no iterative deletions occur and all nodes remain in the $1$-core.
   
The objective and constraints are separable for each $i$ due to $G$ and $G^{soc}$ being directed.\footnote{In the undirected setting, symmetry requires that $G_{ij} = G_{ji}$ and thus all constraints are coupled together.} 
So the planner's problem for activating all agents with $\sum G_{ij} < k_{i}$ decomposes into independent problems. Each has the form $\min\{\sum_{j} \kappa_{ij}\,\left(G^{soc}_{ij} - G_{ij}\right)\}$ subject to $\sum G^{soc}_{ij} \geq  k_{i}$ and the capacity constraint. It is never optimal for the planner to set $\sum_{j} G^{soc}_{ij} > k_{i}$, as this would strictly raise cost without tightening any constraint. Thus $\sum_{j}G^{soc *}_{ij} = k_{i}$. Moreover, the optimal solution adds an outward link to the node $j$ with the cheapest $\{\kappa_{ij}\}$ entry first (which is unique, by assumption). If $k_{i}-G_{ij} > \overline{G}_{ij}$, we proceed to the next cheapest $\kappa_{ij}$. It is clear that the largest value of $\kappa_{ij}$ for which the planner is willing to strengthen $i$'s link to $j$ is given by $\tau_{i}$. 

This linear programming problem is the dual of the fractional knapsack problem, which is well known to be solvable in polynomial time (see \cite[pp.415-417]{korte2011combinatorial}). \hfill \qed

%% file: B.Online_Appendix.tex
\section{Assumptions on utility functions}\label{OA:relaxing_utility}
In this subsection, we formally spell out the assumptions underlying property 1, and discuss what happens when some of the assumptions don't hold. 

We present some sufficient conditions for property 1 to be satisfied
\begin{lem} \label{lem:smoothness_conditions}
    A utility function $u_{i}(x_{i},\sum_{j}G_{ij}x_{j})$ has property 1 if it satisfies
    \begin{enumerate}
        \item \textbf{Complementarity:} $u_{i}(1,\sum_{j}G_{ij}x_{j})-u_{i}(0,\sum_{j}G_{ij}x_{j})$ is strictly increasing in $\sum_{j}G_{ij}x_{j}$
        \item \textbf{Smoothness: }$u_{i}(x_{i},\sum_{j}G_{ij}x_{j})$ is continuous in its second argument. 
        \item \textbf{No Asymptote:} $u_{i}(1,\sum_{j}G_{ij}x_{j})-u_{i}(0,\sum_{j}G_{ij}x_{j})$ is strictly positive for large enough $\sum_{j}G_{ij}x_{j}$, and strictly negative for small enough $\sum_{j}G_{ij}x_{j}$.
    \end{enumerate}
\end{lem}
\begin{proof}
    By intermediate value theorem, smoothness and no asymptote make sure that there exists some $k_{i}$ such that $u_{i}(1,k_{i})-u_{i}(0,k_{i})=0$. By complementarity, for any $k' > k_{i}$, $u_{i}(1,k')-u_{i}(0,k')>0$. For any $k' < k_{i}$, $u_{i}(1,k')-u_{i}(0,k')<0$. 
\end{proof}

The most important assumption is \textbf{complementarity}. Most of our analysis goes through with complementarity. The other two assumptions, and thus property 1, are made only for expositional ease in the main text. We discuss how to perform the analysis without smoothness and no-asymptote. 

\paragraph{Dropping `no-asymptote' condition.} 
If we drop the no-asymptote assumption, then there can exist a set of agents $I_{0}$ such that $u_{i}(1,\sum_{j}G_{ij}x_{j})-u_{i}(0,\sum_{j}G_{ij}x_{j})$ is negative for all $\sum_{j}G_{ij}x_{j}$ and $i \in I_{0}$. There can also exist a set of agents $I_{1}$ such that $u_{i}(1,\sum_{j}G_{ij}x_{j})-u_{i}(0,\sum_{j}G_{ij}x_{j})$ is positive for all $\sum_{j}G_{ij}x_{j}$ and $i \in I_{1}$. 

Agents in $I_{0}$ must always choose action 0 and agents in $I_{1}$ must always choose action 1. In other words, these agents are not strategic -- so we can remove them and adjust the game accordingly. For any agent $i$ not in $I_{0}$ or $I_{1}$, her utility function (in any equilibrium) can be rewritten as
\begin{align*}
    \tilde{u}_{i}(x_{i},\sum_{j \notin I_{0} \cup I_{1} }G_{ij}x_{j} ) \equiv u_{i}(x_{i} , \sum_{j \notin I_{0} \cup I_{1} }G_{ij}x_{j} + \sum_{j \in I_{1}}G_{ij} )
\end{align*}
Since $i \notin I_{0} \cup I_{1}$, $u_{i}(\cdot)$ must satisfy all conditions in \Cref{lem:smoothness_conditions}, which in turn implies that $\tilde{u}_{i}(\cdot)$ also satisfies all these conditions. Therefore, we can just look at the (adjusted) game with the players $N \setminus (I_{0} \cup I_{1})$, who have utility functions $\tilde{u}_{i}$. 

\paragraph{Dropping `smoothness' condition.} If we drop the smoothness condition, then there may not exist a $k_{i}$ such that $u_{i}(1,k_{i})-u_{i}(0,k_{i})=0$. However, given the no-asymptote condition (or when we look at the game after removing the `non-strategic' agents), we can always find some $\hat{k}_{i}$ for each $i$ defined as:
\begin{align*}
    \hat{k}_{i} = \inf \{k:u_{i}(1,k)-u_{i}(0,k) \ge 0 \}.
\end{align*}
We cannot use these $\hat{k}_{i}$ directly as thresholds in the main text. Agent $i$ might strictly prefer action 0 to 1 when $\hat{k}_{i}$ of her neighbours choose action 1. This implies that in the transformed network $H$, agent $i$ might strictly prefer to choose action 0 when exactly 1 of her neighbours is choosing action 1. Therefore, it is possible that agents in the 1-core strictly prefer to choose action 0 even if others in the 1-core are choosing action 1. 

If $u_{i}(1,\hat{k}_{i})-u_{i}(0,\hat{k}_{i}) \ge 0$ for all $i$, then this problem does not arise, and so our analysis (in the main text) goes through by using $\hat{k}_{i}$ as thresholds. However, if this is not the case, then we can perturb the thresholds and then apply our techniques in the main text on these perturbed thresholds. 

By our definition of $\hat{k}_{i}$, for any $\epsilon>0$, we must have $u_{i}(1,\hat{k}_{i}+\epsilon)-u_{i}(0,\hat{k}_{i}+\epsilon) > 0$. This follows from \textbf{complementarity} condition that ensures $u_{i}(1,k)-u_{i}(0,k)$ is \emph{strictly} increasing in $k$. Then, for each $i$ with $u_{i}(1,\hat{k}_{i})-u_{i}(0,\hat{k}_{i}) < 0$, we can find some $\epsilon_{i} > 0$ small enough such that $\sum_{j}G_{ij}x_{j} \notin (\hat{k}_{i}, \hat{k}_{i}+\epsilon]$ for all possible strategy profiles $\mathbf{x}_{-i}$.\footnote{This is because strategy is binary for players and the number of players is finite.} 
It follows from this way of defining $\hat{k}_{i}$ that agent $i$'s best response function can be described as
\begin{align}
    BR_{i}(\mathbf{x}_{-i}) =  \begin{cases}
        0 & \text{if }\sum_{j}G_{ij}x_{j} \le \hat{k}_{i}\\
        1 & \text{otherwise}
    \end{cases}
\end{align}
By how $\epsilon_{i}$ is defined, agent $i$'s best response can also be described by
\begin{align}
    BR_{i}(\mathbf{x}_{-i}) =  \begin{cases}
        1 & \text{if }\sum_{j}G_{ij}x_{j} \ge \hat{k}_{i} + \epsilon_{i} \\
        0 & \text{otherwise}
    \end{cases}
\end{align}
We can then construct a perturbed threshold $\tilde{k}_{i}$ for each agent $i$. If $u_{i}(1,\hat{k}_{i})-u_{i}(0,\hat{k}_{i}) \ge 0$, then just let $\tilde{k}_{i} = \hat{k}_{i}$. If $u_{i}(1,\hat{k}_{i})-u_{i}(0,\hat{k}_{i}) < 0$, then let $\tilde{k}_{i} = \hat{k}_{i} + \epsilon_{i}$. We can then apply our techniques in main text on $(G,\mathbf{\tilde{k}})$ to find the maximal equilibrium. An analogous procedure can be used to characterise the minimal equilibrium. 

\newpage
\section{Transforming to create a fractional threshold game}\label{OA:transform_fractional}
Fractional threshold models are perhaps more common than absolute threshold models, at least in economics. So a natural question is whether we could have transformed the network so as to convert our heterogeneous thresholds into common fractional thresholds. It turns out that the answer is ``\emph{yes, but doing so would not be helpful}''. If we constructed such a transformation, the structure of the transformed network would not be easily relatable to the equilibria of the game.

First, notice that normalising link weights as we do in \Cref{defn:adjustment} would not help here. Because doing so would not affect the fractional threshold.\footnote{If agent $i$ takes action $1$ if and only if $X\%$ of her neighbours (in a weighted sense) do so too, then this remains unaffected if we scale the strength of her links.} 
But adding links to a shadow agent would help. It would change the fraction of links to regular (non-shadow) agents that must be to those who take action $1$ for $i$ to prefer action $1$. But because normalising link weights does not help, we would need \emph{two different shadow agents}: one who non-strategically plays $x_{s1} = 1$ (called S$1$) and another who non-strategically plays $x_{s2} = 0$ (called S$2$). Then we could pick a common fractional threshold -- which we can take as $\frac{1}{2}$ without loss. Finally, for any agent with $q_i > \frac{1}{2}$, create a link of strength $(2q_i - 1) d_i$ to shadow agent S$1$. And for any agent with $q_i < \frac{1}{2}$, create a link of strength $(1 - 2q_i) d_i$ to shadow agent S$2$. Formally,

\begin{defn}[A fractional transformed network, $\mathcal{J}$]
Consider a set of agents $N$ embedded on a network $G$.

\begin{enumerate}
    \item Create two non-strategic ``shadow'' agents, denoted $S1$ and $S2$. Let $x_{S1} = 1$, $x_{S2} = 0$ and let $\mathcal{J}_{S1, i} = \mathcal{J}_{S2, i} = \eta$ for all $i$, some $\eta \geq 1$,

    \item for all $i,j \in N$: let $\mathcal{J}_{ij} = G_{ij}$,

    \item for $i$ with $q_i < \frac{1}{2}$: let $J_{i, S1} = (1 - 2q_i) d_i$ and $\mathcal{J}_{i, S2} = 0$,
    
    \item for $i$ with $q_i > \frac{1}{2}$: let $J_{i, S1} = 0$ and $\mathcal{J}_{i, S2} = (2q_i - 1) d_i$,

    \item  for $i$ with $q_i = \frac{1}{2}$: let $J_{i, S1} = 0$ and $\mathcal{J}_{i, S2} = 0$,
\end{enumerate}
where $d_i = \sum_j G_{ij}$, and $q_i$ is defined in terms of the network $G$.
\end{defn}

It is straightforward to show an analogy to \Cref{lem:adjusted_network} -- that this transformation does not change best responses.

\begin{rem}
    The game with network $\mathcal{J}$ and fractional threshold $\frac{1}{2}$ is best response equivalent to the game $(G, \mathbf{k})$. And best responses are given by:

    \begin{align}\label{eq:BR_fractional}
BR_i(\mathbf{x}_{-i}) =
\begin{cases}
    1 &\text{ if } \quad \sum_{j \in N \cup \{S1, S2\}} \mathcal{J}_{ij} x_j > \frac{1}{2} \sum_{j \in N \cup \{S1, S2\}} \mathcal{J}_{ij}  \\
    \{0,1\} &\text{ if } \quad \sum_{j \in N \cup \{S1, S2\}} \mathcal{J}_{ij} x_j = \frac{1}{2} \sum_{j \in N \cup \{S1, S2\}} \mathcal{J}_{ij}  \\
    0 &\text{ if } \quad \sum_{j \in N \cup \{S1, S2\}} \mathcal{J}_{ij} x_j < \frac{1}{2} \sum_{j \in N \cup \{S1, S2\}} \mathcal{J}_{ij}
\end{cases}
\end{align}
\end{rem}

\begin{proof}
By assumption, $BR_i(\mathbf{x}_{-i}) = 1$ if and only if $\sum_j G_{ij} x_j \geq k_i$.

\noindent \textbf{Case A:} if $q_i < \frac{1}{2}$, then $\mathcal{J}_{i, S1} = (1 - 2q_i)d_i, \mathcal{J}_{i,S2} = 0$. So 
\begin{align*}
     \sum_{j \in N \cup \{S1, S2\}} \mathcal{J}_{ij} x_j &> \frac{1}{2} \sum_{j \in N \cup \{S1, S2\}} \mathcal{J}_{ij} \\
\iff 
     \sum_{j \in N} \mathcal{J}_{ij} x_j + (1 - 2q_i)d_i \cdot 1 + 0 \cdot 0 &> \frac{1}{2} \left(  \sum_{j \in N} \mathcal{J}_{ij} + (1 - 2q_i)d_i + 0 \right) \\
\iff 
     \sum_{j \in N} G_{ij} x_j + (1 - 2q_i)d_i &> \frac{1}{2} \left(  \sum_{j \in N} G_{ij} + (1 - 2q_i)d_i \right) \\
\iff
    \sum_{j \in N} G_{ij} x_j &> \frac{1}{2} \left(  d_i -  d_i + 2q_i d_i \right) \\
\iff 
    \sum_{j \in N} G_{ij} x_j &> q_i d_i
\end{align*}

\noindent \textbf{Case B:} if $q_i > \frac{1}{2}$, then $\mathcal{J}_{i, S1} = 0, \mathcal{J}_{i,S2} = (2q_i - 1)d_i$. So 
\begin{align*}
     \sum_{j \in N \cup \{S1, S2\}} \mathcal{J}_{ij} x_j &> \frac{1}{2} \sum_{j \in N \cup \{S1, S2\}} \mathcal{J}_{ij} \\
\iff 
     \sum_{j \in N} \mathcal{J}_{ij} x_j + 0 \cdot 1 + (2q_i - 1)d_i \cdot 0 &> \frac{1}{2} \left(  \sum_{j \in N} \mathcal{J}_{ij} + 0 + (2q_i - 1)d_i \right) \\
\iff 
     \sum_{j \in N} G_{ij} x_j &> \frac{1}{2} \left(  \sum_{j \in N} G_{ij} + (2q_i - 1)d_i \right) \\
\iff
    \sum_{j \in N} G_{ij} x_j &> q_i d_i
\end{align*}

\noindent \textbf{Case C:} if $q_i = \frac{1}{2}$, then is clear that making no adjustment is sufficient.
\end{proof}

\paragraph{Why this does not work.} The problem here is that once we have transformed the network to make it a game with a constant \emph{fractional} threshold, the cohesiveness of the transformed network no longer corresponds to the equilibria. Notably, once we have a constant fractional threshold, the maximal $\frac{1}{2}$-cohesive set will always be the entire network.\footnote{This is true for any $q \in [0,1]$. And it is clear transforming the network to create a game with a constant fractional threshold $q \notin [0,1]$ will not work.} 
But this clearly does not (in general) correspond to the maximal equilibrium of the original game $G(,\mathbf{k})$.

This problem arises because a pure property of the network structure cannot differentiate between the different behaviours of the two shadow agents. Links to S$1$ and links to S$2$ would affect properties of the network in the same way. But links to S$1$ vs S$2$ have very different implications for equilibrium behaviour. So a $q$-cohesive set of some transformed network with two shadow agents would not correspond to an equilibrium.

The transformation in \Cref{defn:adjustment} worked because there was only a shadow agent taking action $1$ non-strategically. There, we used links from the shadow agent to all others to `force' the shadow agent to be in the $k$-core. Doing that here would only be possible if we had no `never takers' (i.e. those with $q_i > 1$). And having only one shadow agent is insufficient. Shadow agent S$1$ is used to bring some agents `down' to the common threshold. Shadow agent S$2$ is used to bring some agents `up' to the common threshold. 

The heart of the issue is that adding one shadow agent is possible while still having the transformed network structure be meaningful. And one shadow agent can be used to `deal with' either always takers or never takers -- but not both. And the fractional threshold model cannot incorporate either -- and so would need two shadow agents. In contrast, the absolute threshold model automatically accommodates the never takers (who are simply those whose degree is less than the common threshold). So only a single shadow agent is needed to deal with the always takers. 

We need to work through the two shadow agent thought experiment (one shadow who always does 1, other who always does 0) and show that this doesn't work. This is because the straight graph theoretic property of the network cannot deal with different behaviours by the two shadow nodes.

\section{Non-extreme equilibria: dealing with indifferences}\label{OA:non_extreme_indiff} 
In the main text, \Cref{prop:other_eq} provides a characterisation of all equilibria subject to an assumption that no agent is ever completely indifferent. In this section, we drop that `no indifference' assumption. We do so in two ways. 

First, we suppose that thresholds are drawn from some continuous and atomless density function. This makes exactly indifference a probability zero event, as there are only a finite number of threshold values that can make an agent exactly indifferent. It amounts to the same thing as in \Cref{prop:other_eq} -- no indifferences -- but gets there in a somewhat less ad hoc manner.

\begin{prop}\label{prop:other_eq_ext1}
    Suppose $k_i$ is a random draw from some continuous and atomless probability density function $f_i(\cdot)$. Let $M \subseteq N$, and let $H[M]$ be the induced sub-graph of $H(G, \mathbf{k})$. 

    With probability one; for any set $M$, the profile $x_{i} \iff i \in M$ is a Nash Equilibrium if and only if: (i) all agents in $M$ are in the 1-core of $H[M]$, and (ii) $i$ is not in the 1-core of $H[M \cup \{i\}]$ for any $i \in N \setminus M$.
\end{prop}

Second, we allow indifferences. This is the more general version of the result. But it is somewhat harder to interpret. This is because condition (ii) is phrased in terms of a different network. To state the next proposition, recall that $\tilde{\mathbf{k}} =\mathbf{d}- \mathbf{k}$.
\begin{prop}\label{prop:other_eq_ext2}
    Let $M \subseteq N$, and let $H[M]$ be the sub-graph of $H(G, \mathbf{k})$ induced by $M$ and the shadow agent. Similarly let $\tilde{H}[M]$ be the sub-graph of $H(G, \tilde{\mathbf{k}})$ induced by $M$ and the shadow agent. 

    For any set $M$, the profile $x_{i} \iff i \in M$ is a Nash Equilibrium if and only if: (i) All agents in $M$ are in the 1-core of $H[M]$, and (ii) All agents in $N \setminus M$ are in the 1-core of $\tilde{H}[(N \setminus M)]$.
\end{prop}

\subsection{Proofs}
\begin{proof}[\textbf{\emph{Proof of \Cref{prop:other_eq_ext1}}}]
\textbf{Part 1: $H[M]$ is a $1$-core and $H[M \cup \{i\}]$ is \emph{not} a $1$-core $\implies$ $M$ is a Nash Equilibrium.} By the definition of the $1$-core, we have $\sum_{\ell \in M} H_{j \ell} \geq 1$ for all $j \in M$. Therefore all $j \in M$ prefer to choose $x_j = 1$ if all other $j \in M$ do so too (regardless of the choices of agents $i \notin M$). Also by the definition of the $1$-core, we have $\sum_{\ell \in M} H_{i \ell} < 1$ for all $i \notin M$.\footnote{We can see this by a simple contradiction argument. If it were not the case, then $M \cup \{i\}$ would constitute a $1$-core. Which would be a contradiction.} 
So all $i \notin M$ prefer to choose $x_i = 0$ if all other $i \notin M$ do so too (regardless of the choices of agents $j \in M$). Therefore $x_i = 1$ for all $i \in M$ and $x_i = 0$ for all $i \notin M$ is a Nash Equilibrium.

\textbf{Part 2: $M$ is a Nash Equilibrium $\implies$ $H[M]$ is a $1$-core and $H[M \cup \{i\}]$ is \emph{not} a $1$-core.} For $M$ to be a Nash Equilibrium, we must have: $\sum_{\ell} H_{j \ell} x_{\ell} \geq 1$ for all $j \in M$ and $\sum_{\ell} H_{i \ell} x_{\ell} \leq 1$ for all $i \notin M$. Additionally (and by definition), $x_j = 1$ for all $j \in M$ and $x_i = 0$ for all $i \notin M$. Substituting these into the inequalities yields: $\sum_{\ell \in M} H_{j \ell} \geq 1$ for all $j \in M$ and $\sum_{\ell \in M} H_{i \ell} \leq 1$ for all $i \notin M$. 

Now we use the assumption that $k_i$ contains some random component. It means that $Pr(\sum_{\ell \in M} H_{i \ell} = 1) = 0$ for any $B \subseteq N$. This is simply because for any given network $G$ and set $M$, there is a unique value of $k_i$ that yields $\sum_{\ell \in M} H_{i \ell} = 1$. And selecting exactly that value from a continuous distribution is a probability zero event. This means that, with probability one, it is sufficient to consider $\sum_{\ell \in M} H_{i \ell} < 1$ for all $i \notin M$ (rather than the weak inequality version).

Finally, note that by definition having $\sum_{\ell \in M} H_{j \ell} \geq 1$ for all $j \in M$ means $H[M]$ is a $1$-core. And also by definition, having $\sum_{\ell \in M} H_{i \ell} < 1$ for all $i \notin M$ means that $H[M \cup \{i\}]$ is \emph{not} a $1$-core of $H$ for any $i \in N \setminus M$.
\end{proof}

\begin{proof}[\textbf{\emph{Proof of \Cref{prop:other_eq_ext2}}}]
Let $H_{ij}$ denote the $ij$-th element of $H(G,\mathbf{k})$, and $\tilde{H}_{ij}$ denote the $ij$-th element of $H(G,\tilde{\mathbf{k}})$. 
By lemma \ref{lem:adjusted_network}, playing action 1 is a best response for player $i$ iff $\sum_{j \ne i}H_{ij}x_{j} \ge 1 $ and playing action 0 is a best response for $i$ iff $\sum_{j \ne i}\tilde{H}_{ij}(1- x_{j}) \ge 1 $. 

Given that players in set $M$ are choosing action 1 and other players are choosing action 0, player $i$ in set $M$ is playing best responses iff $\sum_{j \ne i, j \in M}H_{ij} \ge 1 $. By definition \ref{defn:k-core}, all agents in $M$ are in 1-core of $H[M]$ iff the set of nodes deleted in the first step is empty, i.e. $\{i \in M: \sum_{j \in M/i} H_{ij} < 1\} = \emptyset$, which is equivalent to $\{i \in M: \sum_{j \in M/i} H_{ij} \ge 1\} = M$. Combining these two facts, we have that players in set $M$ are playing best responses iff they are all in the 1-core of $H[M]$. 

The argument for players in set $N/M$ is similar. Since players in set $M$ are choosing action 1 and other players are choosing action 0, player $i$ in set $N/M$ is playing best responses iff $\sum_{j \ne i, j \in N \setminus M}\tilde{H}_{ij} \ge 1 $. By definition \ref{defn:k-core}, all agents in $N \setminus M$ are in 1-core of $H[N \setminus M]$ iff the set of nodes deleted in the first step is empty, i.e. $\{i \in N \setminus M: \sum_{j\ne i, j \in N \setminus M} \tilde{H}_{ij} < 1\} = \emptyset$, which is equivalent to $\{i \in N \setminus M: \sum_{j \ne i,j \in N \setminus M} \tilde{H}_{ij} \ge 1\} = N \setminus M$. Combining these two facts, we have that players in set $N \setminus M$ are playing best responses iff all agents in $N \setminus M$ are in 1-core of $H[N \setminus M]$. 

The result then follows from definition of Nash equilibrium. 
\end{proof}

\section{Example of changes from the indivisibility}\label{OA:examples}
Here, we present two examples mentioned in \Cref{sec:robustness} of the main text. The second example serves a dual purpose. It shows (a) that moving from continuous to binary actions can dramatically change the key player, and (b) that the impact of a shock can be bounded only by the number of nodes in the network. 

\paragraph{The impact of sequential moves.} Consider a setting with two agents, $i,j$, where $a_i = a_j = 1, \phi_i = \phi_j = 1, G_{ij} = G_{ji} = 1$, $c_i = 1$ and $c_j = 5$ (recall that agents have preferences a la \Cref{eq:BCZ_prefs}). First, suppose agents move simultaneously. It is then easy to verify that when actions are continuous, there is a unique equilibrium with $x_i^* = 1.5, x_j^* = 0.5$. And when actions are binary, we have $k_i = -0.5, k_j = 1.5$ and hence $x_i^* = 1, x_j^* = 0$.

Then suppose that agent $j$ moves before agent $i$. Then it is easy to verify that, when actions are continuous, there is a unique equilibrium with $x_i^* = \frac{5}{3}, x_j^* = \frac{2}{3}$ -- different to the simultaneous move outcome. But when actions are binary, we still have $x_i^* = 1, x_j^* = 0$.

\paragraph{A peripheral key player.} Consider the example in \Cref{fig:inter-centrality}. There are 16 agents, with the network as shown in the figure. All links have weight one, and arrows denote directed links (all others are undirected). Agents $0,1,15$ have thresholds $k_i = 1$, agents $2,3$ have $k_i = 6$, and all others have $k_i = 3$. Additionally, $c_i = 2, \phi_i = 0.1$ for all $i$, and $a_i$ is set to fit the thresholds (recall that $k_i = \frac{1}{\phi_i} (0.5c_i - a_i)$. 

The table then shows the impact of removing each player. The Cascade number is the impact of removing a player in the binary action setting. And \emph{inter-centrality} \citet[Definition 2]{ballester2006} is the impact in the continuous action setting.\footnote{Remember that the binary action setting has multiple equilibria. Here we are considering the impact on the maximal equilibrium. The continuous action setting has a unique equilibrium.} 
With binary actions, agents $0$ and $1$ are the key players. Removing either one of them will cause a catastrophic cascade, resulting in all players switching from action 1 to action 0. In contrast, it is agent $15$ who is the key player when actions are continuous.

\begin{adjustwidth*}{}{-8em}
\begin{figure}[h!]
\centering
\begin{subfigure}[m]{0.45\textwidth}
\hspace{-10mm}
    \resizebox{11.5cm}{!}{ \fbox{
    \input{tikz_pictures/Key_player_peripheral_example} } }
\end{subfigure}
    \qquad
\begin{subfigure}[m]{0.45\textwidth}
\hspace{20mm}
    \resizebox{5.5cm}{!}{ \fbox{
    \input{tikz_pictures/Key_player_table} }}
\end{subfigure}
\caption{All links have weight one, arrows denote directed links. Agents $0,1,15$ have thresholds $k_i = 1$, agents $2,3$ have $k_i = 6$, and all others have $k_i = 3$. We have $c_i = 2, \phi_i = 0.1$ for all $i$, and $a_i$ is set to fit the thresholds (recall $k_i = \frac{1}{\phi_i} (0.5c_i - a_i)$. Inter-centrality shown to 2 decimal places.}\label{fig:inter-centrality}
\end{figure}
\end{adjustwidth*}

This highlights how the move to binary actions can have a very stark impact on who is the key player. 

Additionally, it is clear that weakening the link between agents $0$ and $1$ by any amount (so that the new link strength is $1-\epsilon$ for any $\epsilon \in (0,1]$) will cause agents $0$ and $1$ to both switch to action $0$ in the new maximal equilibrium. In turn, this then causes a catastrophic cascade, resulting in all other agents switching to action $0$ too.

\section{Proof for \Cref{lem:OOC_actions}}\label{OA:cliques_lemma}
We first prove the following claim.
\textbf{Claim:} In any equilibrium, agents in the same overlap must choose the same action. Also, if $x_{i}=1$ and agent $i$ is not in any overlap, then all agents in $i$'s clique must also choose action 1. 
\begin{proof}
    For the first part, suppose that there exists an equilibrium in which it is not true. There are two agents $i$ and $j$ such that they are in the same overlap and $x_{i}=1$, $x_{j}=0$. Note that $i$ and $j$ have the same neighbours, excluding $i$ and $j$ themselves. Therefore, the set of $i$'s neighbours who choose action 1 are also neighbours of agent $j$. In addition, $i$ chooses action 1. Therefore, $j$ has more neighbours who choose action 1 than agent $i$. Agent $j$ has incentive to deviate to choose action 1. 

    For the second part, suppose not. Again, there are two agents $i$ and $j$ such that they are in the same clique and $x_{i}=1$, $x_{j}=0$. We know that the set of agent $j$'s neighbours must be a superset of the set of agent $i$'s neighbours. the rest follows in essentially the same way as last paragraph.
\end{proof}

Consider any equilibrium and let $X^{1}$ denote the set of agents choosing action 1 in that equilibrium. By the claim, for any $i$, we know that if $i$ is in some overlap $B_{l}$, then $B_{l} \subseteq X^{1}$. If $i$ is in no overlap and is in clique $C_{l}$, then $\Gamma_{l} \subseteq X^{1}$. We can then find an overlap or a clique associated with each $i \in X^{1}$. The argument above shows that the union of these overlaps and cliques is a subset of $X^{1}$.
It's trivial that the union of these overlaps and cliques is also a superset of $X^{1}$ by how these overlaps and cliques are defined. 

\section{Comparative Statics}\label{OA:comp_statics}
\begin{rem}\label{rem:comp stat}
\phantom{2}

\noindent (i) \ \ $x_i^{MIN}$ and $x_i^{MAX}$ are weakly increasing in $G_{j \ell}$ and weakly decreasing in $k_j$, for all $i,j,\ell$. 

\noindent (ii) \ if the game exhibits \emph{positive spillovers}, $u_i^{MIN}$ and $u_i^{MAX}$ (which denote agent $i$'s payoff in the minimal and maximal equilibrium respectively) are weakly increasing in $G_{j \ell}$ and weakly decreasing in $k_j$, for all $i,j,\ell$.
\end{rem}

\begin{proof}
Follows from \citet[Theorem 13]{milgrom1994monotone}.
\end{proof}

\section{A Planner's Problem}\label{OA:planner_problem}
Here, we consider an alternative social planner's problem: one where the planner aims to maximise adoption of the action subject to a budget constraint. We show that this problem, in general, NP-hard. It has a tractable solution in the very restrictive case where the planner starts from an empty network. As before, we assume that the planner can induce agents to coordinate on the maximal Nash equilibrium. This is the equilibrium that the planner most prefers. And if actions have positive utility spillovers, it is also the equilibrium that agents most prefer. 

Additionally, we give the social planner an extra tool -- we allow her to alter thresholds (as well as links). We consider a non-transferable utility setting, where the planner has a budget $B > 0$ she can `spend' on modifying both thresholds and links. 
And to try and aid simpler results, we assume that the cost of building links is a constant. Specifically, the marginal cost of lowering a threshold is $1$ (by a normalisation) and the marginal cost of strengthening a link is $\omega > 0$. The following definition sets this out formally.

\begin{defn}\label{def:planners_problem}
Planner's problem: $\max_{(\Delta G, \mathbf{\Delta k})} \sum_{i \in N} x_i^{MAX}$, where $\Delta G_{i} \geq 0$ and $\Delta k_{i} \leq 0$ $\forall i$ and is subject to $-\sum_{i} \Delta k_{i} + \omega\sum_{i} \sum_{j}\Delta G_{ij} \leq B$,\footnote{In an abuse of notation, we are reusing $B$ to refer to the planner's budget here. This is different to it use in \Cref{sec:non-extreme eqa}, where we used it for the set of bridges.} 
where $u_i = \sum_j G_{ij} x_i x_j - k_i x_i$, the $^{`MAX'}$ superscript denotes the maximal Nash Equilibrium, and $\alpha$ is the degree of inequality aversion.
\end{defn}
 
It turns out that in general, this problem is computationally hard. Specifically, it is \emph{NP-hard}. This means that the solution to the problem we set out in \Cref{def:planners_problem} will not always be solvable within a useful amount of time. \cite{sipser2006introduction} pp. 275-304 provides an introduction to NP-hardness and its implications.

\begin{prop}\label{prop:NP_hard}
   The adoption planner's and the utility planner's problems are both NP-hard.
\end{prop}

This result is driven by the fact that strengthening even a single link or lowering a single threshold can have wider impacts on the structure of the $1$-core of the adjusted network $H$. The proof involves showing that the budget-constrained planners' problems are each equivalent to the maximum coverage problem -- which is  known to be NP-hard \citep{khuller1999budgeted}. We largely follow the strategy of \cite{zhou2019k} for showing that the link-addition problem is NP hard, but must accommodate the more fine-grained interventions available to our social planner. Both the link-addition and threshold-subsidy problems have been studied in the computer science literature (where they are typically referred to as `edge-addition' and `vertex anchoring', respectively). Work there has shown that any polynomial time approximation of these problems will perform \emph{arbitrarily poorly} -- they are inapproximable.\footnote{\cite{downey2013fundamentals} and \cite{vazirani2001approximation} provide discussions on inapproximability.} 

So not only is the problem hard to solve, but we cannot even guarantee a good \emph{approximate} solution within a useful amount of time \citep{zhang2017finding, zhou2019k}. These problems have been studied on unweighted and undirected networks, but we might expect the generalisation to unweighted and undirected networks to be even more computationally intensive and thus the same approximation problems would remain.

Before proving \Cref{prop:NP_hard}, it is helpful to first state and prove the following lemma.

\begin{lem}\label{lem:strat_equiv}
    Consider the game $(G,\mathbf{k})$, and suppose that $k_i < 0$ for some $i$. For any changes to the thresholds $\mathbf{\Delta  k} \in \mathbb{R}_{-}^{n}$, there exists changes to the link weights $\Delta G  \in \mathbb{R_{+}}^{n\times n}$ such that the game $(G + \Delta G, \mathbf{k})$ has the same maximal equilibrium as the game $(G,\mathbf{k} + \mathbf{\Delta  k})$. The converse for $\Delta G  \in \mathbb{R}_{+}^{n\times n}$ is also true.
\end{lem}

\paragraph{Proof of \Cref{lem:strat_equiv}} It is clear that two games have the same maximal equilibria if their transformed graphs have the same agents in their $1$-core. If $k_{i}+\Delta k_{i}>0$, then $i$ is in the the $1$-core of $H$ iff $\frac{\sum _{j \in C(H,1)}\Delta G_{ij}}{k_{i}+\Delta k_{i}} \geq 1$,\footnote{We know $C(H,1)$ is non-empty by assumption} so we may simply choose any $\Delta G$ satisfying $\sum_{j \in C(H,1)} \Delta G_{ij} = -\Delta k_i$.  If $k_{i}+\Delta k_{i}<0$, we set $\Delta G_{ij} = -\Delta k_i$ for some $j \in K^{-}$.  To see the converse statement, let $H'$ be the transformed graph of $G+\Delta G$, and simply set $\Delta k_{i} = \frac{\sum_{j \in C(H')}G_ij}{\sum_{j \in C(H')}G_{ij}+\Delta G_{ij}}$. \hfill\qed

\paragraph{Proof of \Cref{prop:NP_hard}} We show NP hardness for the edge addition-problem as \Cref{lem:strat_equiv} immediately implies the NP-hardness result for the threshold design problem. 

The proof largely follows from \cite{zhou2019k}, which shows that the budget-constrained $k$-core optimisation problem is NP-hard on unweighted and un-directed graphs for $k \geq 3$. However, we allow our planner to make weighted and directed edge additions instead. We therefore adapt their proof in what follows below for our purposes.

First, note that in any weighted graph where all edges have of weight $1/k$ ($k \in \mathbb{R}^{+}$), then extending the $1$-core is equivalent to extending the $k$-core in an unweighted graph. 
We therefore drop the lower bound on $k$ in our result above for the NP-hardness result on weighted graphs.

We will show that the problem of maximising the adoption of the action on a graph with positive integer threshold $k \geq 3$\footnote{Note that although we allow $k$ to be an arbitrary real number, for the proof of NP-hardness it is sufficient to hardness for the positive integer $k$ instance of the problem.} for taking the action by adding at most $B$ weighted links at cost $\omega$ can be reduced to the unweighted budget constrained maximum coverage problem.

The inputs to the maximum coverage problem are a collection of sets $T_1, T_2, \ldots T_{c}$, and a budget $B$. We have that $\cup_{1 \leq i \leq c} T_{i}$ consists of $d$ elements $\{e_1 \ldots e_{d}\}$\footnote{The sets may have non-empty intersection.} Each set has an associated cost $C(T_i) \in \mathbb{R}^{+}$. The aim is to choose sets from this collection which cover the maximum number of elements $e_{i}$ subject to the budget constraint. The maximum coverage problem is known to be NP-hard.  We suppose that $c>k$, and $\sum_{i=1}^{c} C(T_{i})>B$. We also suppose that each of the chosen sets contains at least two elements without loss of generality.

For a given instance of the budget-constrained maximum coverage problem, we shall construct a graph $G$ on which the $k$-core maximisation problem is equivalent. The graph we construct is unweighted an undirected, but it is isomorphic to a weighted and directed graph.
The vertices of the constructed graph lie in one of three parts: $M$, $N$, and $Q$. The first part, $M$, consists of $c$ sets of $d+3$ vertices each, indexed $u_{j}^{i}$. Thus $M = \cup_{1 \leq i \leq c} M_i$, where $M_{i} = \{u_{j}^{i} | 1\leq j \leq d+3\}$. $N$ consists of $d$ vertices indexed $v_i$. $Q$ consists of a $k+1$ clique. 

For every set $M_i$, we create a cycle by connecting $u_{j}^{i}$ to $u_{j+1}^{i}$ for $1\leq j \leq d+2$, and connecting $u_1^{i}$ to $u_{d+3}^{i}$.  We connect the vertices in $N$ to the vertices in $M$ by adding a link between $v_{i}$ and $u_{i}^{j}$ if $e_{i} \in T_{j}$.  We then connect the vertices in $M _i\setminus \{u_{d+1}^{i},u_{d+3}^{i}\}$ which have degree less than $k$ to the vertices in $Q$ so that each vertex has degree $k$. We then add edges between $\{u_{d+1}^{i},u_{d+3}^{i}\}$ and Q so that each of the two vertices has degree $k-1$. We finally add $k-1$ edges between each of the vertices in $N$ and the vertices in $Q$. 

The $k$-core algorithm will delete all the vertices except those in $Q$. The most cost-effective solution to expand the $k$-core  is to add edges of weight 1 between $u^{i}_{d+1}$ and $u^{i}_{d+3}$, which will add all of $M_{i}$ to the $k$-core, as well as and $v_j$ which are adjacent to $M_{i}$ (which corresponds exactly with the set of nodes in $T_{i}$). Adding edges of weight larger or smaller than 1 is not cost-efficient as any node in the graph needs an out-degree increment of exactly 1 to be added. 

Another solution would be to connect nodes in $M_i$ and $M_j$. However, an equivalent number of agents can be added to the $k$-core by instead connecting $u^{i}_{d+1}$ and $u^{i}_{d+3}$ (as $M_i$ can only be added to the $k$-core if all of its nodes are added). Adding an edge between any vertices in N is not worthwhile, as we assume that any chosen set has at least two elements - and thus adding any $M_{i}$ to the $k$-core will add at least two vertices in $N$ to the k-core.

Thus if there is a polynomial-time solution to the budget-constrained edge $k$-core problem, there is a solution to the budget-constrained maximum coverage problem.

%% file: tikz_pictures/Key_player_peripheral_example.tex
\begin{tikzpicture}[scale=1, 
  every node/.style={circle, draw, minimum size=2cm, font=\huge}, 
  every path/.style={thick}
]

\tikzset{
  mid arrow/.style={
    decoration={markings, mark=at position 0.6 with {\arrow[scale=3]{latex}}},
    postaction={decorate} } }
\node (n0) at (12,0) {0};
\node (n1) at (12,4) {1};

\node (n15) at (0,0) {15};

\def \R {6}  

\foreach \i in {0,...,12} {
  \pgfmathsetmacro{\angle}{360/13*\i}
  \pgfmathsetmacro{\x}{\R*cos(\angle)}
  \pgfmathsetmacro{\y}{\R*sin(\angle)}
  \pgfmathsetmacro{\cx}{0}
  \pgfmathsetmacro{\cy}{0}
  \pgfmathsetmacro{\px}{\cx + \x}
  \pgfmathsetmacro{\py}{\cy + \y}
  \pgfmathtruncatemacro{\n}{\i + 2}
  \node (n\n) at (\px,\py) {\n};
}

  \draw (n2) -- (n15);
  \draw (n3) -- (n15);
  \draw (n4) -- (n15);
  \draw (n5) -- (n15);
  \draw (n6) -- (n15);
  \draw (n7) -- (n15);
  \draw (n8) -- (n15);
  \draw (n9) -- (n15);
  \draw (n10) -- (n15);
  \draw (n11) -- (n15);
  \draw (n12) -- (n15);
  \draw (n13) -- (n15);
  \draw (n14) -- (n15);
  \draw (n0) to (n1);
  \draw[mid arrow] (n2) to (n0);
  \draw[mid arrow] (n3) to (n0);  
  \draw[mid arrow] (n2) to (n1);
  \draw[mid arrow] (n3) to (n1);
  \draw (n2) -- (n3);
  \draw (n3) -- (n4);
  \draw (n4) -- (n5);
  \draw (n5) -- (n6);
  \draw (n6) -- (n7);
  \draw (n7) -- (n8);
  \draw (n8) -- (n9);
  \draw (n9) -- (n10);
  \draw (n10) -- (n11);
  \draw (n11) -- (n12);
  \draw (n12) -- (n13);
  \draw (n13) -- (n14);
  \draw (n14) -- (n2);
  \draw[bend left=45] (n2) to (n4);
  \draw[bend left=45] (n3) to (n5);
  \draw[bend left=45] (n4) to (n6);
  \draw[bend left=45] (n5) to (n7);
  \draw[bend left=45] (n6) to (n8);
  \draw[bend left=45] (n7) to (n9);
  \draw[bend left=45] (n8) to (n10);
  \draw[bend left=45] (n9) to (n11);
  \draw[bend left=45] (n10) to (n12);
  \draw[bend left=45] (n11) to (n13);
  \draw[bend left=45] (n12) to (n14);
  \draw[bend left=45] (n13) to (n2);
  \draw[bend left=45] (n14) to (n3);
\end{tikzpicture}

%% file: tikz_pictures/Key_player_table.tex
\begin{tabular}{ccc}
\hline
Agent & \begin{tabular}[c]{@{}c@{}}Cascade \\ Number \end{tabular} & \begin{tabular}[c]{@{}c@{}}Inter-\\ centrality\end{tabular}  \\
\hline
0 & 15 & 1.01 \\
1 & 15 & 1.01 \\
2 & 0 & 2.63  \\
3 & 0 & 2.63 \\
4 & 0 & 2.95 \\
5 & 0 & 2.99  \\
6 & 0 & 3.03 \\
7 & 0 & 3.04 \\
8 & 0 & 3.04 \\
9 & 0 & 3.04 \\
10 & 0 & 3.04 \\
11 & 0 & 3.04 \\
12 & 0 & 3.03 \\
13 & 0 & 2.99 \\
14 & 0 & 2.95 \\
15 & 0 & 11.82
\end{tabular}

%% file: main.bbl
\begin{thebibliography}{58}
\providecommand{\natexlab}[1]{#1}
\providecommand{\url}[1]{\texttt{#1}}
\expandafter\ifx\csname urlstyle\endcsname\relax
  \providecommand{\doi}[1]{doi: #1}\else
  \providecommand{\doi}{doi: \begingroup \urlstyle{rm}\Url}\fi

\bibitem[Abello and Queyroi(2013)]{abello2013fixed}
J.~Abello and F.~Queyroi.
\newblock Fixed points of graph peeling.
\newblock In \emph{Proceedings of the 2013 IEEE/ACM International Conference on Advances in Social Networks Analysis and Mining}, pages 256--263, 2013.

\bibitem[Acemoglu et~al.(2011)Acemoglu, Ozdaglar, and Yildiz]{acemoglu2011diffusion}
D.~Acemoglu, A.~Ozdaglar, and E.~Yildiz.
\newblock Diffusion of innovations in social networks.
\newblock In \emph{2011 50th IEEE conference on decision and control and European control conference}, pages 2329--2334. IEEE, 2011.

\bibitem[Akbarpour et~al.(2020)Akbarpour, Malladi, and Saberi]{akbarpour2020just}
M.~Akbarpour, S.~Malladi, and A.~Saberi.
\newblock Just a few seeds more: value of network information for diffusion.
\newblock \emph{Available at SSRN 3062830}, 2020.

\bibitem[Alexander et~al.(2022)Alexander, Forastiere, Gupta, and Christakis]{alexander2022algorithms}
M.~Alexander, L.~Forastiere, S.~Gupta, and N.~A. Christakis.
\newblock Algorithms for seeding social networks can enhance the adoption of a public health intervention in urban india.
\newblock \emph{Proceedings of the National Academy of Sciences}, 119\penalty0 (30):\penalty0 e2120742119, 2022.

\bibitem[Allmis et~al.(2025)Allmis, Elliott, Ghiglino, and Langtry]{allmis2025social}
P.~Allmis, M.~Elliott, C.~Ghiglino, and A.~Langtry.
\newblock Mobile but miserable.
\newblock \emph{Working Paper}, 2025.

\bibitem[Aral et~al.(2013)Aral, Muchnik, and Sundararajan]{aral2013engineering}
S.~Aral, L.~Muchnik, and A.~Sundararajan.
\newblock Engineering social contagions: Optimal network seeding in the presence of homophily.
\newblock \emph{Network Science}, 1\penalty0 (2):\penalty0 125--153, 2013.

\bibitem[Ballester et~al.(2006)Ballester, Calv{\'o}-Armengol, and Zenou]{ballester2006}
C.~Ballester, A.~Calv{\'o}-Armengol, and Y.~Zenou.
\newblock Who's who in networks. wanted: The key player.
\newblock \emph{Econometrica}, 74\penalty0 (5):\penalty0 1403--1417, 2006.

\bibitem[Banerjee et~al.(2013)Banerjee, Chandrasekhar, Duflo, and Jackson]{banerjee2013diffusion}
A.~Banerjee, A.~G. Chandrasekhar, E.~Duflo, and M.~O. Jackson.
\newblock The diffusion of microfinance.
\newblock \emph{Science}, 341\penalty0 (6144):\penalty0 1236498, 2013.

\bibitem[Bass(1969)]{bass1969new}
F.~M. Bass.
\newblock A new product growth for model consumer durables.
\newblock \emph{Management science}, 15\penalty0 (5):\penalty0 215--227, 1969.

\bibitem[Bernheim et~al.(1987)Bernheim, Peleg, and Whinston]{bernheim1987coalition}
B.~D. Bernheim, B.~Peleg, and M.~D. Whinston.
\newblock Coalition-proof nash equilibria i. concepts.
\newblock \emph{Journal of economic theory}, 42\penalty0 (1):\penalty0 1--12, 1987.

\bibitem[Bollob{\'a}s(1984)]{bollobas1984evolution}
B.~Bollob{\'a}s.
\newblock The evolution of sparse graphs.
\newblock \emph{Graph theory and combinatorics (Cambridge, 1983)}, pages 35--57, 1984.

\bibitem[Bonacich(1987)]{bonacich1987power}
P.~Bonacich.
\newblock Power and centrality: A family of measures.
\newblock \emph{American Journal of Sociology}, 92\penalty0 (5):\penalty0 1170--1182, 1987.

\bibitem[Bramoull{\'e} and Kranton(2016)]{bramoulle2016games}
Y.~Bramoull{\'e} and R.~Kranton.
\newblock Games played on networks.
\newblock \emph{Working Paper}, 2016.

\bibitem[Centola and Macy(2007)]{centola2007complex}
D.~Centola and M.~Macy.
\newblock Complex contagions and the weakness of long ties.
\newblock \emph{American Journal of Sociology}, 113\penalty0 (3):\penalty0 702--734, 2007.

\bibitem[Chwe(1999)]{chwe1999structure}
M.~S.-Y. Chwe.
\newblock Structure and strategy in collective action.
\newblock \emph{American journal of sociology}, 105\penalty0 (1):\penalty0 128--156, 1999.

\bibitem[Chwe(2000)]{chwe2000communication}
M.~S.-Y. Chwe.
\newblock Communication and coordination in social networks.
\newblock \emph{The Review of Economic Studies}, 67\penalty0 (1):\penalty0 1--16, 2000.

\bibitem[Downey and Fellows(2013)]{downey2013fundamentals}
R.~G. Downey and M.~R. Fellows.
\newblock \emph{Fundamentals of Parameterized Complexity}.
\newblock Springer, New York, 2013.
\newblock ISBN 978-1-4471-5558-4.
\newblock \doi{10.1007/978-1-4471-5559-1}.

\bibitem[Easley and Kleinberg(2010)]{easley2010networks}
D.~Easley and J.~Kleinberg.
\newblock \emph{Networks, crowds, and markets: Reasoning about a highly connected world}.
\newblock Cambridge university press, 2010.

\bibitem[Gagnon and Goyal(2017)]{gagnon2017networks}
J.~Gagnon and S.~Goyal.
\newblock Networks, markets, and inequality.
\newblock \emph{American Economic Review}, 107\penalty0 (1):\penalty0 1--30, 2017.

\bibitem[Galeotti et~al.(2010)Galeotti, Goyal, Jackson, Vega-Redondo, and Yariv]{galeotti2010network}
A.~Galeotti, S.~Goyal, M.~O. Jackson, F.~Vega-Redondo, and L.~Yariv.
\newblock Network games.
\newblock \emph{The review of economic studies}, 77\penalty0 (1):\penalty0 218--244, 2010.

\bibitem[Ghiglino and Tabasso(2024)]{ghiglino2024endogenous}
C.~Ghiglino and N.~Tabasso.
\newblock Endogenous identity in a social network.
\newblock \emph{arXiv preprint arXiv:2406.10972}, 2024.

\bibitem[Goyal(2023)]{goyal2023networks}
S.~Goyal.
\newblock \emph{Networks: An economics approach}.
\newblock MIT Press, 2023.

\bibitem[Granovetter(1978)]{granovetter1978threshold}
M.~Granovetter.
\newblock Threshold models of collective behavior.
\newblock \emph{American journal of sociology}, 83\penalty0 (6):\penalty0 1420--1443, 1978.

\bibitem[Granovetter and Soong(1986)]{granovetter1986threshold}
M.~Granovetter and R.~Soong.
\newblock Threshold models of interpersonal effects in consumer demand.
\newblock \emph{Journal of Economic Behavior \& Organization}, 7\penalty0 (1):\penalty0 83--99, 1986.

\bibitem[Guilbeault et~al.(2018)Guilbeault, Becker, and Centola]{guilbeault2018complex}
D.~Guilbeault, J.~Becker, and D.~Centola.
\newblock Complex contagions: A decade in review.
\newblock \emph{Complex spreading phenomena in social systems: Influence and contagion in real-world social networks}, pages 3--25, 2018.

\bibitem[Harkins(2013)]{harkins2013network}
A.~Harkins.
\newblock Network games with perfect complements.
\newblock \emph{Warwick University Draft, unpublished}, 2013.

\bibitem[Hellmann(2013)]{hellmann2013existence}
T.~Hellmann.
\newblock On the existence and uniqueness of pairwise stable networks.
\newblock \emph{International Journal of Game Theory}, 42:\penalty0 211--237, 2013.

\bibitem[Hellmann(2021)]{hellmann2021pairwise}
T.~Hellmann.
\newblock Pairwise stable networks in homogeneous societies with weak link externalities.
\newblock \emph{European Journal of Operational Research}, 291\penalty0 (3):\penalty0 1164--1179, 2021.

\bibitem[Hiller(2017)]{hiller2017peer}
T.~Hiller.
\newblock Peer effects in endogenous networks.
\newblock \emph{Games and Economic Behavior}, 105:\penalty0 349--367, 2017.

\bibitem[Jackson(2008)]{jackson2008}
M.~O. Jackson.
\newblock \emph{Social and economic networks}.
\newblock Princeton University Press, 2008.

\bibitem[Jackson and Storms(forthcoming)]{jackson2025behavioral}
M.~O. Jackson and E.~C. Storms.
\newblock Behavioral communities and the atomic structure of networks.
\newblock \emph{American Economic Journal: Microeconomics}, forthcoming.

\bibitem[Jackson and Yariv(2006)]{jackson2006diffusion}
M.~O. Jackson and L.~Yariv.
\newblock Diffusion on social networks.
\newblock \emph{Economie publique/Public economics}, 1\penalty0 (16), 2006.

\bibitem[Jackson and Zenou(2015)]{jackson2015games}
M.~O. Jackson and Y.~Zenou.
\newblock Games on networks.
\newblock In \emph{Handbook of game theory with economic applications}, volume~4, pages 95--163. Elsevier, 2015.

\bibitem[Karp(2009)]{karp2009reducibility}
R.~M. Karp.
\newblock Reducibility among combinatorial problems.
\newblock In \emph{50 Years of Integer Programming 1958-2008: from the Early Years to the State-of-the-Art}, pages 219--241. Springer, 2009.

\bibitem[Kempe et~al.(2003)Kempe, Kleinberg, and Tardos]{kempe2003maximizing}
D.~Kempe, J.~Kleinberg, and {\'E}.~Tardos.
\newblock Maximizing the spread of influence through a social network.
\newblock In \emph{Proceedings of the ninth ACM SIGKDD international conference on Knowledge discovery and data mining}, pages 137--146, 2003.

\bibitem[Kempe et~al.(2005)Kempe, Kleinberg, and Tardos]{kempe2005influential}
D.~Kempe, J.~Kleinberg, and {\'E}.~Tardos.
\newblock Influential nodes in a diffusion model for social networks.
\newblock In \emph{international colloquium on automata, languages, and programming}, pages 1127--1138. Springer, 2005.

\bibitem[Khuller et~al.(1999)Khuller, Moss, and Naor]{khuller1999budgeted}
S.~Khuller, A.~Moss, and J.~S. Naor.
\newblock The budgeted maximum coverage problem.
\newblock \emph{Information processing letters}, 70\penalty0 (1):\penalty0 39--45, 1999.

\bibitem[Kong et~al.(2019)Kong, Shi, Wu, and Zhang]{kong2019k}
Y.-X. Kong, G.-Y. Shi, R.-J. Wu, and Y.-C. Zhang.
\newblock k-core: Theories and applications.
\newblock \emph{Physics Reports}, 832:\penalty0 1--32, 2019.

\bibitem[Korte et~al.(2011)Korte, Vygen, Korte, and Vygen]{korte2011combinatorial}
B.~H. Korte, J.~Vygen, B.~Korte, and J.~Vygen.
\newblock \emph{Combinatorial optimization}, volume~1.
\newblock Springer, 2011.

\bibitem[Leister et~al.(2022)Leister, Zenou, and Zhou]{leister2022social}
C.~M. Leister, Y.~Zenou, and J.~Zhou.
\newblock Social connectedness and local contagion.
\newblock \emph{The Review of Economic Studies}, 89\penalty0 (1):\penalty0 372--410, 2022.

\bibitem[Malliaros et~al.(2020)Malliaros, Giatsidis, Papadopoulos, and Vazirgiannis]{malliaros2020core}
F.~D. Malliaros, C.~Giatsidis, A.~N. Papadopoulos, and M.~Vazirgiannis.
\newblock The core decomposition of networks: Theory, algorithms and applications.
\newblock \emph{The VLDB Journal}, 29\penalty0 (1):\penalty0 61--92, 2020.

\bibitem[Milgrom and Roberts(1990)]{milgrom1990rationalizability}
P.~Milgrom and J.~Roberts.
\newblock Rationalizability, learning, and equilibrium in games with strategic complementarities.
\newblock \emph{Econometrica: Journal of the Econometric Society}, pages 1255--1277, 1990.

\bibitem[Milgrom and Shannon(1994)]{milgrom1994monotone}
P.~Milgrom and C.~Shannon.
\newblock Monotone comparative statics.
\newblock \emph{Econometrica: Journal of the Econometric Society}, pages 157--180, 1994.

\bibitem[Morris(2000)]{morris2000contagion}
S.~Morris.
\newblock Contagion.
\newblock \emph{The Review of Economic Studies}, 67\penalty0 (1):\penalty0 57--78, 2000.

\bibitem[Morris and Ui(2004)]{morris2004best}
S.~Morris and T.~Ui.
\newblock Best response equivalence.
\newblock \emph{Games and Economic Behavior}, 49\penalty0 (2):\penalty0 260--287, 2004.

\bibitem[Pastor-Satorras et~al.(2015)Pastor-Satorras, Castellano, Van~Mieghem, and Vespignani]{pastor2015epidemic}
R.~Pastor-Satorras, C.~Castellano, P.~Van~Mieghem, and A.~Vespignani.
\newblock Epidemic processes in complex networks.
\newblock \emph{Reviews of modern physics}, 87\penalty0 (3):\penalty0 925, 2015.

\bibitem[Reich(2023)]{reich2023diffusion}
B.~Reich.
\newblock The architecture of social networks and the diffusion of innovations.
\newblock \emph{Working Paper}, 2023.

\bibitem[Rodrigues(2018)]{rodrigues2018network}
F.~A. Rodrigues.
\newblock Network centrality: an introduction.
\newblock In \emph{A mathematical modeling approach from nonlinear dynamics to complex systems}, pages 177--196. Springer, 2018.

\bibitem[Sadler and Golub(2021)]{sadler2021games}
E.~Sadler and B.~Golub.
\newblock Games on endogenous networks.
\newblock \emph{arXiv preprint arXiv:2102.01587}, 2021.

\bibitem[Saxena and Iyengar(2020)]{saxena2020centrality}
A.~Saxena and S.~Iyengar.
\newblock Centrality measures in complex networks: A survey.
\newblock \emph{arXiv preprint arXiv:2011.07190}, 2020.

\bibitem[Schelling(1978)]{schelling2006micromotives}
T.~C. Schelling.
\newblock \emph{Micromotives and macrobehavior}.
\newblock WW Norton \& Company, 1978.

\bibitem[Seidman(1983)]{seidman1983network}
S.~B. Seidman.
\newblock Network structure and minimum degree.
\newblock \emph{Social networks}, 5\penalty0 (3):\penalty0 269--287, 1983.

\bibitem[Sipser(2006)]{sipser2006introduction}
M.~Sipser.
\newblock \emph{Introduction to the Theory of Computation}.
\newblock Course Technology, second edition, 2006.
\newblock ISBN 7111173279 9787111173274.

\bibitem[Vazirani(2001)]{vazirani2001approximation}
V.~V. Vazirani.
\newblock \emph{Approximation Algorithms}.
\newblock Springer, Berlin, Heidelberg, 2001.
\newblock ISBN 978-3-540-65367-7.
\newblock \doi{10.1007/978-3-662-04565-7}.

\bibitem[Vives(1990)]{vives1990nash}
X.~Vives.
\newblock Nash equilibrium with strategic complementarities.
\newblock \emph{Journal of Mathematical Economics}, 19\penalty0 (3):\penalty0 305--321, 1990.

\bibitem[Zenou and Zhou(2022)]{zenou2022network}
Y.~Zenou and J.~Zhou.
\newblock Network games made simple.
\newblock \emph{Available at SSRN 4225140}, 2022.

\bibitem[Zhang et~al.(2017)Zhang, Zhang, Qin, Zhang, and Lin]{zhang2017finding}
F.~Zhang, Y.~Zhang, L.~Qin, W.~Zhang, and X.~Lin.
\newblock Finding critical users for social network engagement: The collapsed k-core problem.
\newblock In \emph{Thirty-First AAAI Conference on Artificial Intelligence}, 2017.

\bibitem[Zhou et~al.(2019)Zhou, Zhang, Lin, Zhang, and Chen]{zhou2019k}
Z.~Zhou, F.~Zhang, X.~Lin, W.~Zhang, and C.~Chen.
\newblock K-core maximization: An edge addition approach.
\newblock In \emph{IJCAI}, pages 4867--4873, 2019.

\end{thebibliography}
